# Research and experimental verification on low-frequency long-range sound propagation characteristics under ice-covered and range-dependent marine environment in the Arctic

Jinbao Weng[1,a], Yubo Qi[2], Yanming Yang[1], Hongtao Wen[1], Hongtao Zhou[1], Ruichao Xue[1]

1. Laboratory of Ocean acoustics and Remote Sensing, Institute of Oceanography, Ministry of Natural Resources, Xiamen, Fujian 361005, China
2. State key laboratory of acoustics, Institute of Acoustics, Chinese Academy of Sciences, Beijing 100190, China

## ABSTRACT

At present, research on sound propagation under the Arctic ice mainly focuses on modeling and experimental verification of sound propagation under sea ice cover and unique sound velocity profiles. Among them, the main research object of concern is sound transmission loss, and this article will delve into the time-domain waveform and fine dispersion structure of low-frequency broadband acoustic signals. Firstly, based on the theory of normal modes, this article derives the horizontal wavenumber expression and warping transformation operator for refractive normal modes in the Arctic deep-sea environment. Subsequently, based on measured ocean environmental parameters and sound field simulation calculations, this article studied the general laws of low-frequency long-range sound propagation signals in the Arctic deep-sea environment, and elucidated the impact mechanism of environmental factors such as seabed terrain changes, horizontal changes in sound velocity profiles (SSPs), and sea ice cover on low-frequency long-range sound propagation in the Arctic. This article validates the above research viewpoint through a sound propagation experiment conducted in the Arctic with a propagation distance exceeding 1000km. The marine environment of this experiment has obvious horizontal variation characteristics. At the same time, this article takes the lead in utilizing the warping transformation of refractive normal waves in the Arctic waters to achieve single hydrophone based separation of normal waves and extraction of dispersion structures, which is conducive to future research on underwater sound source localization and environmental parameter inversion based on dispersion structures.

## I. INTRODUCTION

The study of underwater sound propagation in the Arctic began in the 1950s and 1960s. Lepage and Schmidt conducted sound field modeling research on sound propagation under the Arctic ice. The marine environment of the Arctic has undergone significant changes in recent decades. In addition to the warming of the Arctic and the reduction of sea ice caused by global warming, the Canadian

---

[a] Email: wengjinbao@tio.org.cn

Basin has shown significant dual-channel sound velocity profiles, known as Beaufort lenses. Recently, Americans conducted a one-year sound propagation experiment in the Canadian Basin to observe the impact of changes in the Arctic marine environment on sound propagation. They analyzed the changes in the SSP of the Beaufort Sea over a one-year period, and also analyzed the impact of changes in the Pacific summer water layer on sound propagation in the Beaufort Sea. At the same time, changes in sea ice and SSPs were used to explain the phenomenon of sound propagation loss fluctuations reaching 60 dB. In addition, the Americans conducted sound propagation experiments at different frequencies in the Canadian Basin using nuclear-powered submarines in the ICEX-18 experiment, and explained the reasons for the changes in propagation losses using the vertical distribution of eigenfunctions of normal modes of different numbers and frequencies.

Remote underwater acoustic propagation in the Arctic is also used to observe the Arctic marine environment, including studying changes in Arctic seawater temperature. Multiple related experiments have been conducted, with the main purpose of achieving acoustic tomography of the Arctic Ocean. At present, low-frequency single frequency acoustic signals are mainly used in the acoustic tomography work of the Arctic Ocean, and the amount of information obtained through this method is very limited. For example, the ACOUS project team used a single frequency signal of 20.5Hz to emit acoustic signals on Franz Josef Island from October 1998 to December 1999, which were received by a vertical array in the Lincon Sea 1250km away, and by a receiving device at Barros point 2720km away. In 1994, researchers conducted a similar TAP experiment. At the same time, Americans measured the SSP of seawater crossing the Arctic Ocean in three experiments: SCICE-95, SCICEX-98, and SCICEX-99. In 2019, Europe and North America jointly conducted the CAATEX experiment, which focused on how much the Arctic Ocean had changed over more than two decades. The experimental research mentioned above mainly utilizes low-frequency single frequency acoustic signals.

At present, research on low-frequency long-range sound propagation in the deep Arctic mainly uses low-frequency single frequency signals, and then uses large aperture synchronous vertical arrays to carry out normal mode separation work. For low-frequency broadband pulse signals, T C Yang strictly derived the eigenvalues and dispersion structures of normal modes under typical Arctic SSPs. Based on the dispersion structure of a single normal mode, the sound source localization was achieved, and a large aperture synchronous vertical array based sound source localization work was also carried out.

In the past decade, more and more research has been conducted on the separation of normal modes using warping transform. At present, warping transformation is mainly used in shallow sea areas, and various warping transformation operators in shallow sea marine environments are considered, including horizontally invariant marine environments and horizontally changing marine environments. At the same time, research on time-domain warping transformation and frequency-domain Warping transformation has been carried out. Also, time-domain and frequency-domain signal autocorrelation warping transformations have been carried out, and various modifications have been made to warping transformations. Based on this, research has been carried out on modal separation, dispersion structure extraction, ocean environment inversion, target positioning, and other aspects in various marine environments.

Julien et al. utilized the low-frequency broadband acoustic signals collected from whales in the Arctic waters, and separated multiple normal modes after removing the influence of the sound source spectrum, thereby achieving the localization of whale sound sources. The current research mainly utilizes the warping transformation of shallow sea reflected normal modes. Qi et al. have derived the warping transformation of shallow sea refractive normal modes, but there is currently no derivation work on the warping transformation of refractive normal modes in Arctic deep-sea waveguides, nor has there been any subsequent extraction of dispersion curves and separation of normal modes. For the common airgun sound signals in Arctic waters, Thode et al. analyzed the sound propagation loss of low-frequency long-distance airgun sound signals, and Keen et al. analyzed the dispersion structure of airgun sound signals in shallow and deep Arctic waters, as well as the variation of the intensity of airgun sound signals with propagation distance. However, they did not conduct modal separation of signals and did not delve into the impact of Arctic marine environment on sound propagation.

Section II of this article is based on the theory of normal modes, deducing the approximate expression of the horizontal wavenumber of the normal mode in the Arctic surface seawater waveguide, and deriving the normal mode warping transformation operator under this waveguide. The effectiveness of the approximate expression is verified through simulation. Section III describes the use of simulation to study the time-domain waveform and dispersion structure of acoustic signals under two typical SSP conditions in the Arctic, and analyzes their formation mechanism using eigenfunctions. Section IV provides a detailed introduction to a low-frequency long-range sound propagation experiment under Arctic ice, especially the seabed topography, seawater SSP, and sea ice coverage of the experimental sea area. Section V uses the data from this Arctic sound propagation experiment to validate the research viewpoints obtained from simulation, especially the impact of those three environmental factors on the characteristics of sound signals. At the same time, a single hydrophone normal mode separation method based on warping transform was implemented using experimental data under typical SSP conditions in the Arctic.

## II. THEORYS

### A. Normal mode theory

According to the book of << Computational Ocean Acoustics >>, the ocean sound field can be represented by normal modes, and the sound pressure at the receiving point can be represented by a series of superposition of normal modes,

$$P(\omega,r,z) = S(\omega) \frac{je^{-j\pi/4}}{\rho(z_s)\sqrt{8\pi r}} \sum_{m=1}^{M} \Psi_m(z_s) \Psi_m(z_r) \frac{e^{-\alpha_m(\omega)r} e^{jk_{rm}(\omega)r}}{\sqrt{k_{rm}(\omega)}}. \quad (1)$$

Among them, $\omega$ represents the angular frequency, $k_{rm}(\omega)$ represents the horizontal wave number, $\Psi_m(z)$ represents the eigenfunction, $z_s$ and $z_r$ represents the source depth and the receiver depth, $r$ represents the distance between source and receiver, $M$ represents the number of effective normal modes, $S(\omega)$ represents the spectral level of the sound source. Assuming the spectral level of a broadband pulse sound source is 1, the sound signal pressure can be abbreviated as follows,

$$P(\omega,r,z) = \sum_{m=1}^{M} A_m(\omega) e^{jk_{rm}(\omega)r}. \quad (2)$$

Among them, $A_m(\omega) = \frac{e^{j\pi/4} \Psi_m(z_s) \Psi_m(z_r) e^{-\alpha_m(\omega)r}}{\rho(z_s)\sqrt{8\pi r k_{rm}(\omega)}}$.

## B. Approximate expression for horizontal wavenumber of surface seawater waveguides in the Arctic

Qi et al. derived the horizontal wavenumber of refractive normal modes in shallow water waveguides and derived the corresponding warping transformation operator. In the typical deep sea SSP environment of the Arctic, the sound velocity of the surface 400 meters of water increases approximately linearly with depth, forming an refractive normal mode acoustic waveguide. When the energy of refractive normal modes is mainly concentrated at the surface 400 meters, this article deduces an approximate expression for the horizontal wave number in this case. The SSP of the surface 400 meters of water can be approximated as shown in equation (3),

$$c(z) = c_0(1+az) \quad (3)$$

Among them, $c_0$ is the sound velocity on the surface of seawater, $a$ is the rate at which the sound velocity increases with depth, $z$ is the depth of seawater, and $c(z)$ is the sound velocity of seawater at depth $z$. According to the WKB approximation, the horizontal wave number of a normal mode can be obtained by solving the following equation,

$$\phi(k_{rm},\omega) + \Delta\phi_{dn}(k_{rm},\omega) + \Delta\phi_{up}(k_{rm},\omega) = 2(m-1)\pi, m=1,2,3... \quad (4)$$

Among them, $\phi(k_{rm},\omega)$ is the phase shift of a normal mode propagating through a span of water,

$$\phi(k_{rm},\omega)=2\int_{z_1}^{z_2}k_z(z,k_{rm})dz=2\int_{z_1}^{z_2}\sqrt{\left(\frac{\omega}{c(z)}\right)^2-k_{rm}^2}dz \qquad (5)$$

And $z_1$ is the sea surface depth or the depth of upward reversal point, $z_2$ is the seabed depth or the depth of downward reversal point. For the surface waveguide formed by 400 meters of water in the Arctic surface, for the convenience of derivation, it is assumed that the sea surface is not covered by sea ice, and the interface between seawater and air is an ideal interface, with a reflection phase shift of $\Delta\phi_{up}(k_{rm},\omega)=-\pi$; the phase shift of the refractive normal mode at the lower reversal point is $\Delta\phi_{dn}(k_{rm},\omega)=-\pi/2$. By substituting these three terms into formula (4), the following expression can be obtained,

$$\int_0^\varsigma \sqrt{\left(\frac{\omega}{c_0(1+az)}\right)^2-k_{rm}^2}dz=\left(m-\frac{1}{4}\right)\pi, m=1,2,3... \qquad (6)$$

Among them, $\varsigma$ is the depth of the downward reversal point. After derivation, the left term of equation (6) can be approximated as follows,

$$\int_0^\varsigma \sqrt{\left(\frac{\omega}{c_0(1+az)}\right)^2-k_{rm}^2}dz \approx \int_0^\varsigma \sqrt{k_0^2(1-2az)-k_{rm}^2}dz \qquad (7)$$

Among them, $k_0=\omega/c_0$. By combining equations (6) and (7), the following can be derived,

$$k_{rm}=\frac{\omega}{c_0}\sqrt{1-b_{1m}\omega^{-\frac{2}{3}}} \qquad (8)$$

$$b_{1m}=\left[3a\pi\left(m-\frac{1}{4}\right)c_0\right]^{\frac{2}{3}} \qquad (9)$$

By derivation, an approximate expression for the horizontal wave number of a normal mode can be obtained as follows,

$$k_{rm}\approx \frac{\omega}{c_0}-\frac{1}{2}\frac{b_{1m}}{c_0}\omega^{\frac{1}{3}} \qquad (10)$$

Next, simulation calculations will be conducted using the measured SSP of the central ice region of the Arctic to verify the correctness of formula (10). Figure 1 shows the measured SSP in the central ice region of the Arctic. The SSP measurement time was August 5, 2017, and the coordinates of the measurement station were (179.6°E, 80.0°N), and the maximum depth measured was 1666m.

Linear approximation was performed on this SSP. The sea surface sound velocity is approximately 1434m/s, the water sound velocity is approximately 1459m/s at a depth of 400m, and the water sound velocity is approximately 1473m/s at a depth of 1660m. From this figure, it can be seen that the approximate linear SSP is basically consistent with the measured SSP. Using the sound field model KrakenC to calculate the horizontal wave numbers of the top ten normal modes under a linear SSP, and comparing them with the calculation results of formula (10), it can be seen from the figure that for the top seven normal modes, the two are basically the same, but as the normal mode index increases, the difference between the two also increases. According to the subsequent simulation of the normal wave eigenfunction, it can be concluded that as the index of normal modes increases, the vertical coverage depth of the eigenfunction also gradually increases. When the coverage depth exceeds 400m, the premise assumption for deriving formula (10) is that energy is concentrated at the surface 400m, which is not valid, resulting in significant calculation errors in the approximate formula.

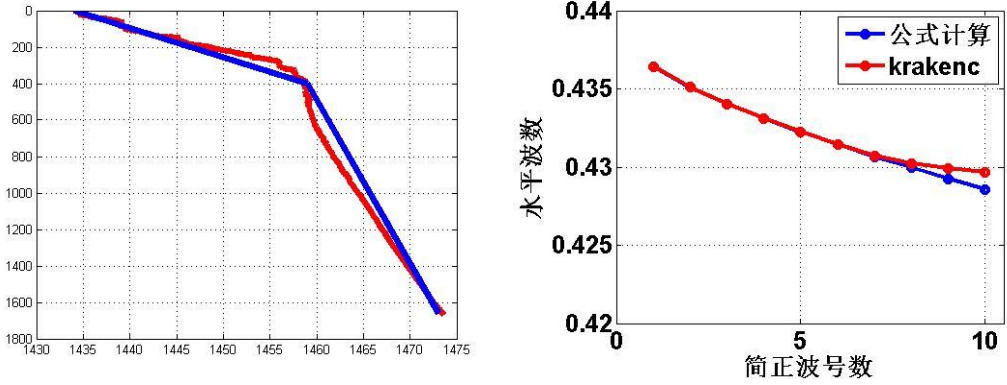

Figure 1 Comparison between the linear SSP obtained by approximating the measured SSP, and the comparison between the horizontal wave number estimated based on the approximate formula and the model simulation results.

## C. Warping transformation operator for refractive normal modes in Arctic surface seawater waveguides

For surface waveguides in typical deep sea environments in the Arctic, the frequency domain expression for refracted normal modes is as follows,

$$P(\omega, r, z) = \sum_{m=1}^{M} A_m(\omega) e^{j\left(\omega t_r - \frac{1}{2} t_r b_{1m} \omega^{1/3}\right)} \quad (11)$$

Among them, the arrival time $t_r$ of the acoustic signal is

$$t_r = \frac{r}{c_0} \quad (12)$$

After derivation, the expression of refractive normal modes in the time domain is

$$P(t,r,z) = \frac{1}{2\pi} \sum_{m=1}^{M} \int_{-\infty}^{\infty} A_m(\omega) e^{j\left[\omega(t-t_r) + \frac{1}{2}t_r b_{1m}\omega^{1/3}\right]} d\omega \quad (13)$$

For a pulsed sound source, the amplitude of the normal mode wave slowly changes with frequency. Based on the stable phase method, the following can be approximated,

$$P(t,r,z) = \frac{1}{2\pi} \sum_{m=1}^{M} A_m(\omega_{ms}) \sqrt{\frac{2\pi}{|\varphi_m''(\omega_{ms},t)|}} e^{j\left[\omega_{ms}(t-t_r) + \frac{1}{2}t_r b_{1m}\omega_{ms}^{1/3} + \frac{\pi}{4}\text{sgn}(\varphi_m''(\omega_{ms},t))\right]} \quad (14)$$

Among them, sgn represents a symbolic function.

$$\varphi_m(\omega,t) = \omega(t-t_r) + \frac{1}{2}t_r b_{1m}\omega^{1/3} \quad (15)$$

Among them, $\varphi_m''(\omega,t)$ represents the second derivative of $\varphi(\omega,t)$ with respect to diagonal frequency. The stable phase point satisfies the following equation,

$$(t-t_r) + \frac{1}{6}t_r b_{1m}\omega^{1/3} = 0 \quad (16)$$

By solving the equation, the stable phase angle frequency expression can be obtained as,

$$\omega_{ms} = \left(\frac{6(t_r - t)}{t_r b_{1m}}\right)^{-3/2} \quad (17)$$

By substituting equation (17) into the above equation, the expression for the instantaneous phase of the normal mode can be obtained as follows:

$$\varphi_m(\omega_{ms},t) = \omega_{ms}(t-t_r) + \frac{1}{2}t_r b_{1m}\omega_{ms}^{1/3} = (t_r-t)^{-1/2}(t_r b_{1m})^{3/2} 2^{-1/2} 3^{-3/2} \quad (18)$$

Therefore, the warping transformation operator in the time domain can be obtained as follows,

$$h(t) = t_r - t^{-2} \quad (19)$$

The corresponding inverse operator is as follows,

$$h^{-1}(t) = (t_r - t)^{-1/2} \quad (20)$$

The frequency of the normal mode after warping transformation is as follows,

$$f_m = r^{3/2} c_0^{-1/2} 2^{-3/2} 3^{-1/2} \left[a\left(m - \frac{1}{4}\right)\right] \quad (21)$$

From formula (21), it can be seen that the transformed normal mode frequency is related to the sound propagation distance, sea surface sound velocity, sound velocity gradient, and normal mode index.

# III. SIMULATION OF SOUND FIELD CHARACTERISTICS

In the past few decades, the marine environment of the Arctic has undergone significant changes, especially in the Canadian Basin and the Chukchi Plateau, where obvious dual-channel SSPs have emerged. Figure 2 shows a comparison of the SSPs of the same sea area obtained from historical data in the literature in 1994 and 2016, which clearly shows this change. The changes in the SSP of Arctic seawater have attracted the interest of underwater acoustic researchers, especially the advantage of dual-channel SSPs in facilitating the long-distance propagation of sound signals. Currently, multiple large-scale experiments have been conducted to study the sound field effects of dual-channel SSPs.

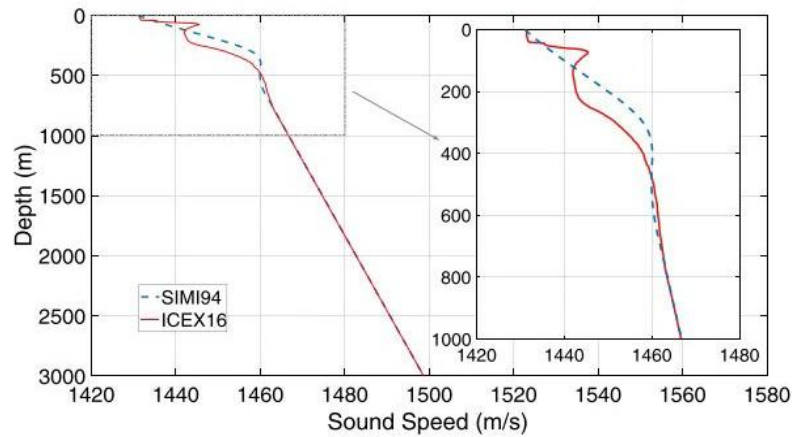

FIG. 1. (Color online) Sound speed profile during SIMI94 (dashed) and ICEX16 (solid).

Figure 2 Comparison of seawater SSPs in the same Arctic waters from historical data in the literature in 1994 and 2016

This article mainly focuses on the sound propagation issues in waters such as the Arctic Pacific sector and the central ice region. Common SSPs in these waters include the following two types: typical SSPs in the central ice region, and dual-channel SSPs in the Canadian Basin and Chukchi Plateau. Figure 3 shows two types of SSPs measured in the central ice region and the Chukchi Plateau in this sound propagation experiment. It can be seen from the figure that the main difference between these two types of SSPs is the SSP of the surface water about 400m, while the SSP of water over 400m is basically consistent and monotonically increasing with depth.

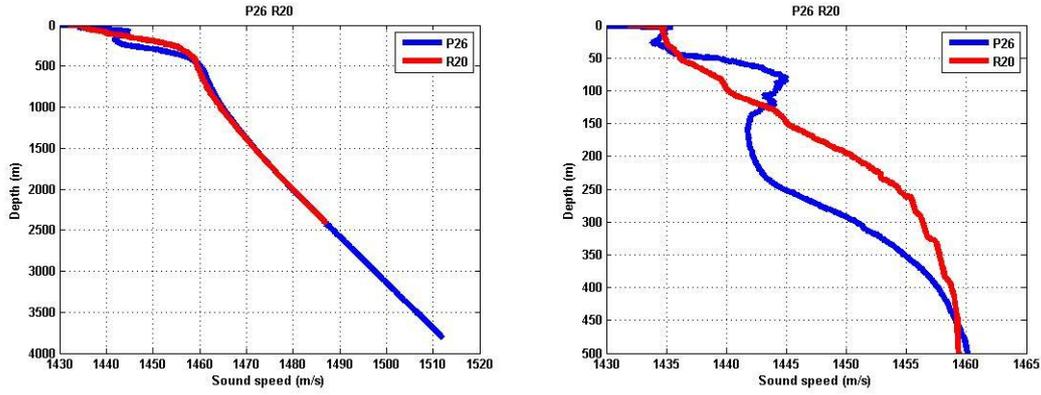

Figure 3 Comparison of SSPs measured in the Arctic central ice region and the Chukchi Plateau during sound propagation experiments

Next, this article is based on the Kraken model for calculating the normal mode sound field, conducting sound field simulations in typical deep sea SSPs in the Arctic and dual-channel SSPs in typical Chukchi Plateau environments, including two-dimensional sound transmission loss, time-domain waveform and dispersion structure of sound signals, and normal mode eigenfunctions. During the simulation process, in order to maintain consistency with the experiment as much as possible, it is assumed that the sound source depth is 300m, the sound source frequency bandwidth is from 10Hz to 100Hz, and the seawater depth is 3800m.

## A. Simulation of two-dimensional sound transmission loss

Based on the Kraken model for calculating the normal mode sound field, two-dimensional sound transmission loss was calculated under two typical deep sea SSPs in the Arctic, assuming a sound source frequency of 100Hz. Figure 4 shows the calculated two-dimensional sound transmission loss. From the figure, it can be seen that the sound propagation loss under two different SSPs has a significant difference in the surface 400 m water body, and there is not much difference between the two in the water greater than 400 m, indicating that the difference in the surface SSP mainly affects the spatial distribution of the surface sound field.

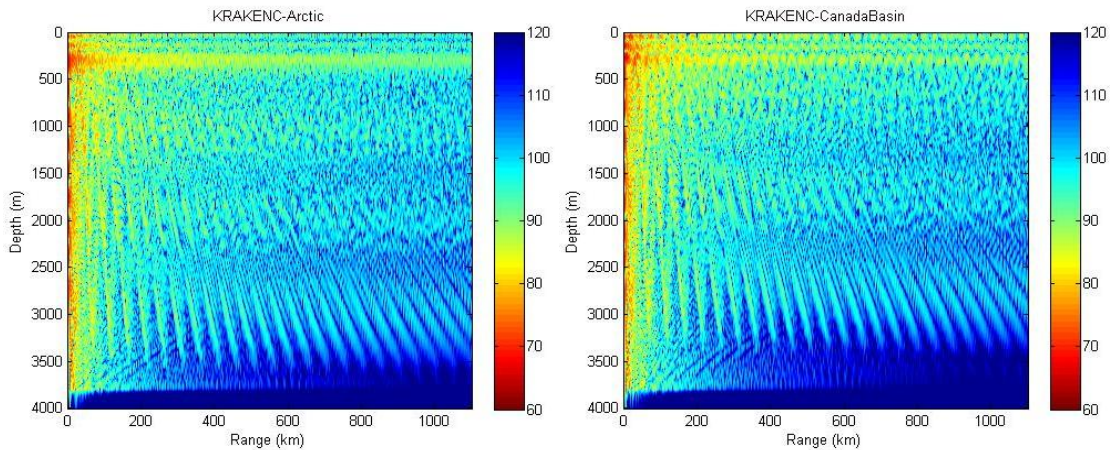

Figure 4 Simulation results of two-dimensional sound transmission loss in typical deep sea

## B. Simulation of time-domain waveform and dispersion structure of acoustic signals

Next, this article conducts simulation analysis of surface received acoustic signals at different distances under typical deep sea SSP in the Arctic, including time-domain waveforms and dispersion structures. In order to maintain consistency with the experimental settings as much as possible, it is assumed that the propagation distances of sound signals are 105km, 199km, 864km, and 1066km, respectively. At the same time, the dispersion structure is calculated using the normal mode model Kraken, and compared with the time-frequency analysis of simulated sound signals. Then, the time-domain waveform and dispersion structure of sound signals will be analyzed, as well as the impact of changes in SSP on both.

From Figure 5, it can be seen that under the SSP condition of the typical central ice region in the Arctic, the time-domain waveform of the received sound signal consists of two parts, including multipath sound ray arrival without dispersion characteristics and multimodal arrival with dispersion characteristics. The former has a large horizontal grazing angle, which covers most of the sea water, thus having a higher sound velocity and arriving earlier; the latter is mainly influenced by the normal wave waveguide effect formed by surface sea water, thus having a dispersion structure, and the latter arrives later due to the lower sound velocity in surface sea water. At the same time, the duration of the acoustic signal increases with distance, due to the stable sound velocity difference between the front and rear parts of the acoustic signal. In addition, in the time-frequency analysis of simulated acoustic signals, the dispersion structure of each normal mode does not fully cover the dispersion curve calculated by the model, indicating that the normal mode has a clear upper frequency limit.

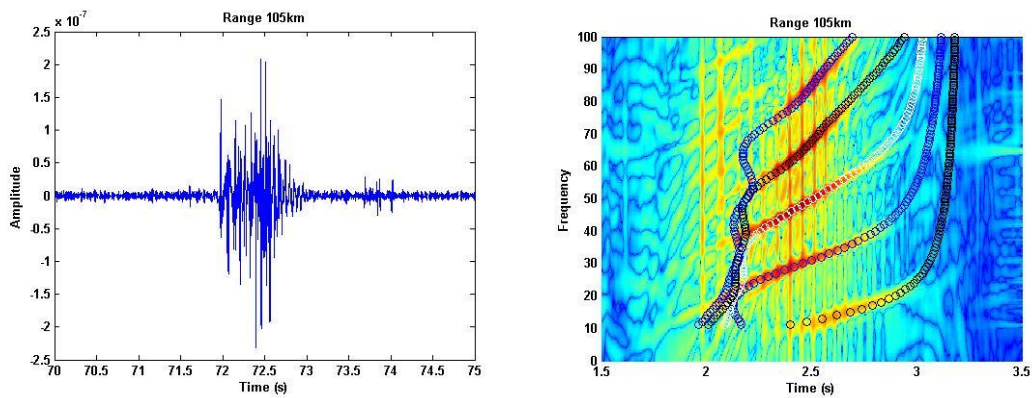

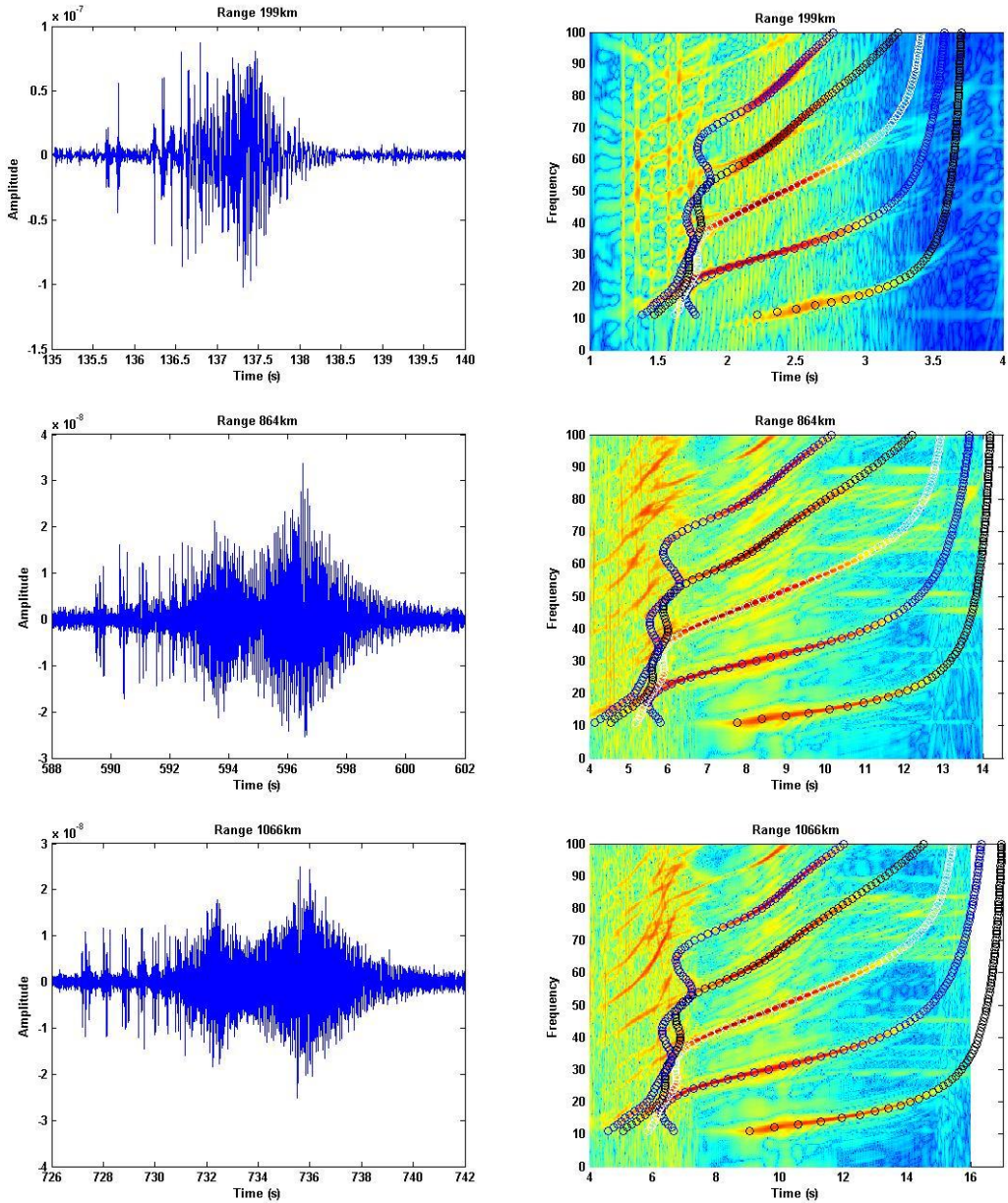

Figure 5 Time domain waveform and dispersion structure of acoustic signals at different distances simulated under typical deep sea SSP conditions in the Arctic

Figure 6 shows the time-domain waveform and dispersion structure of the acoustic signal under typical dual-channel SSP conditions. It can be seen from the figure that the time-domain waveform and dispersion structure under dual-channel SSP are different from the simulation results under typical deep sea SSP in the Arctic. In terms of time-domain waveform, the sound signal under the dual-channel SSP lasts longer, especially at ultra long distances; In terms of dispersion structure, the dispersion structure of the sound signal under the dual-channel SSP is relatively complex and chaotic, and there is a phenomenon of crossover in the dispersion curve, especially at higher frequencies, which is inconsistent with the common pattern of lower order normal waves arriving later. At some frequencies, the first normal wave arrives earlier than the second normal wave.

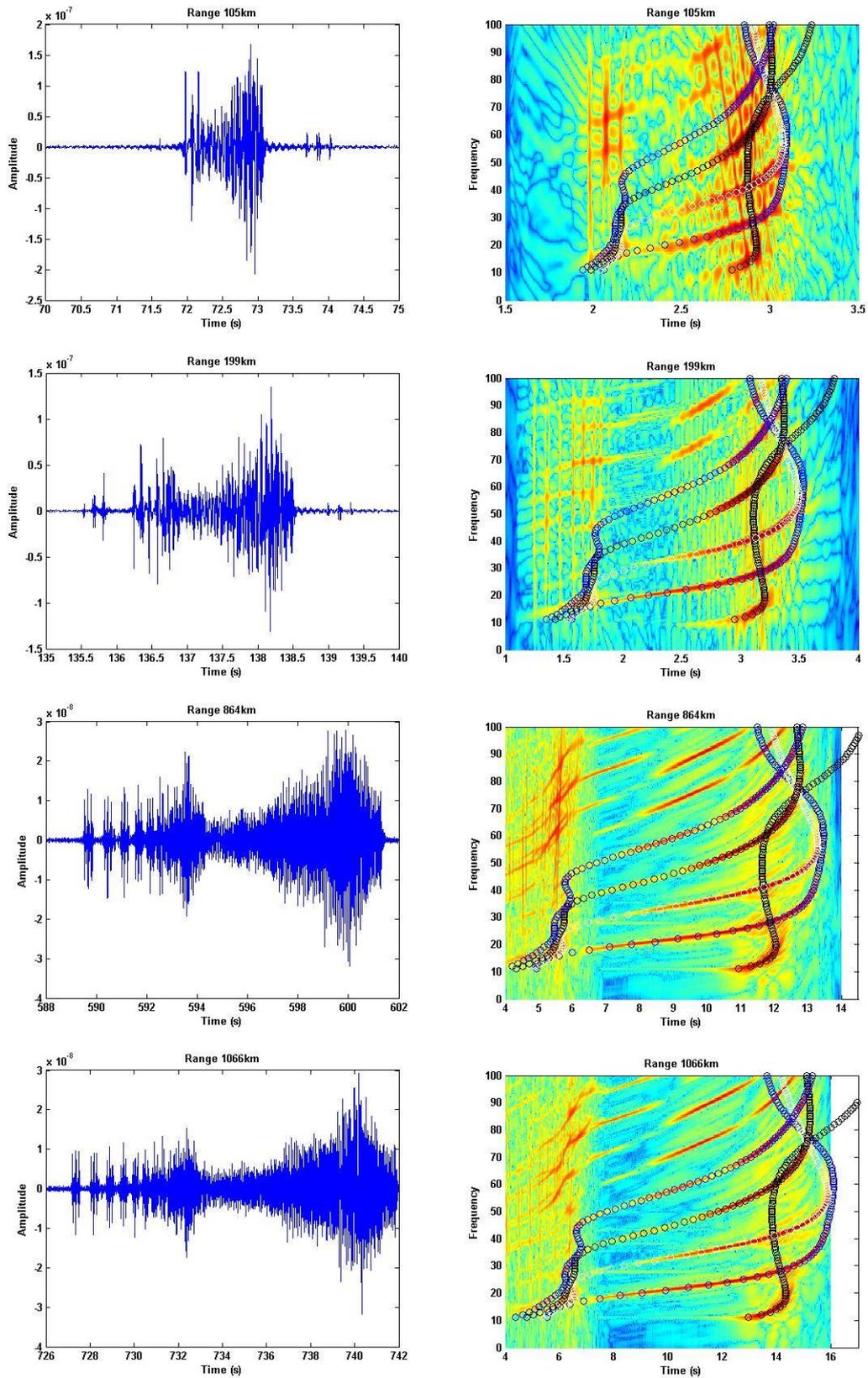

Figure 6 Time domain waveform and dispersion structure of acoustic signals at different distances simulated under typical dual-channel SSP conditions in the Arctic

## C. Simulation and analysis of eigenfunctions of normal modes

Next, this article conducts simulation analysis of the vertical distribution of normal wave eigenfunctions at different frequencies under typical deep sea SSP in the Arctic. Due to limited space, this article only selected four typical frequencies for analysis, including 11Hz, 40Hz, 70Hz, and 100Hz, with a depth range of 0m to 1000m. From Figure 7, it can be seen that under the typical deep sea SSP conditions in the Arctic, the energy distribution range of each normal mode eigenfunction gradually accumulates into the surface sea water as the frequency increases. Due to the increase of sound velocity with depth in the typical Arctic SSP, the low-frequency part of each normal mode has a larger group velocity than the high-frequency part, thus arriving earlier, which is consistent with the results in the simulation. When the sound source or receiver has a large working depth, the normal mode at higher frequencies cannot be excited or received, resulting in an upper limit on the frequency of the received sound signal, which is consistent with the results in simulation. At the same time, for the lower frequency part of each normal mode, the eigenfunction covered the greater depth part of sea water, making the normal mode more susceptible to the influence of the seabed. Therefore, the lower frequency limit of normal modes is mainly constrained by the seabed, and this phenomenon can be observed in experimental data.

In addition, it can be seen from the figure that the higher the frequency of a certain number of normal modes, the more concentrated the energy distribution of the eigenfunction in the surface seawater. Therefore, when there is sea ice cover on the sea surface, the higher the frequency of normal modes, the more severe the influence of sea ice and the faster the attenuation. Overall, due to the vertical distribution characteristics of the eigenfunctions of normal waves in the unique marine environment of the Arctic, the low-frequency part of each normal wave arrives earlier than the high-frequency part. Each normal wave has an upper and lower limit frequency, which is affected by the depth of the sound source, reception depth, and sea ice cover, while the lower limit frequency is affected by the seabed.

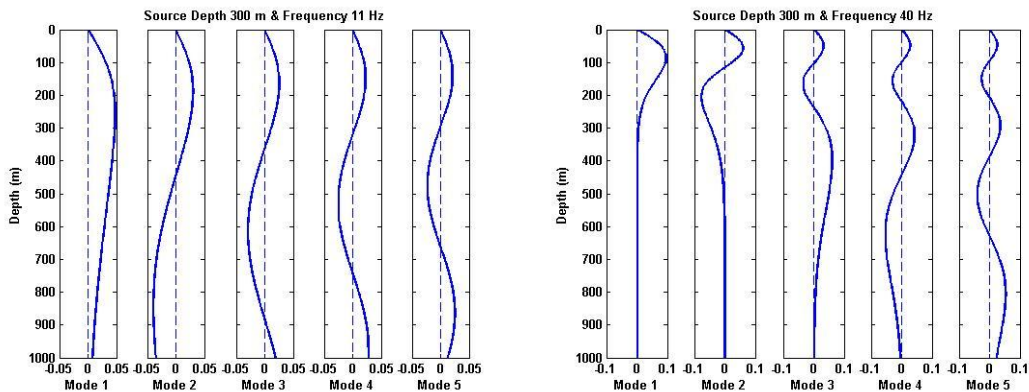

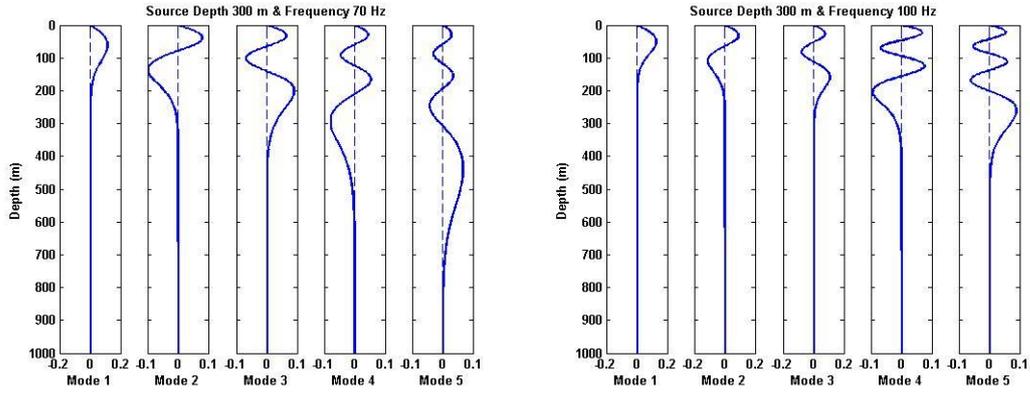

Figure 7 Vertical distribution of normal mode eigenfunctions with different frequencies simulated under typical deep sea SSP conditions in the Arctic

Figure 8 shows the vertical distribution of normal wave eigenfunctions at different frequencies under typical dual-channel SSP conditions. It can be seen from the figure that the vertical distribution of normal mode eigenfunctions has changed under dual-channel SSP conditions, especially when the frequency is high and the energy of the eigenfunctions is mainly concentrated in the surface water. Due to the fact that the sound velocity of the surface water under the dual-channel SSP first increases, then decreases, and then increases with depth, this leads to a crossover phenomenon in the dispersion structure, especially in the higher frequency parts. Taking 40Hz as an example, the eigenfunction of the first normal mode covers the water body with faster sound velocity, which results in the first normal mode arriving earlier than the second normal mode. In addition, due to the smaller overall sound velocity of the surface water under the condition of dual-channel SSP, the high-frequency part of each mode wave arrives later, resulting in an increase in the time difference between low-frequency and high-frequency arrival, and a longer duration of the time domain waveform, which is consistent with the results in simulation.

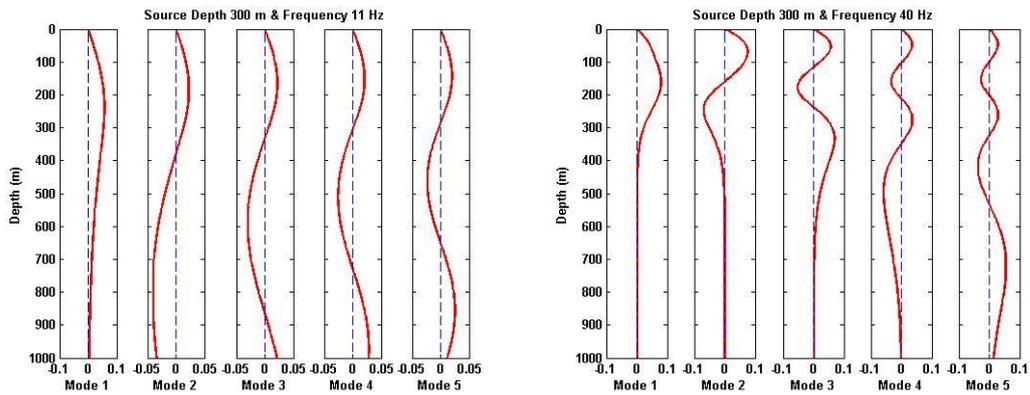

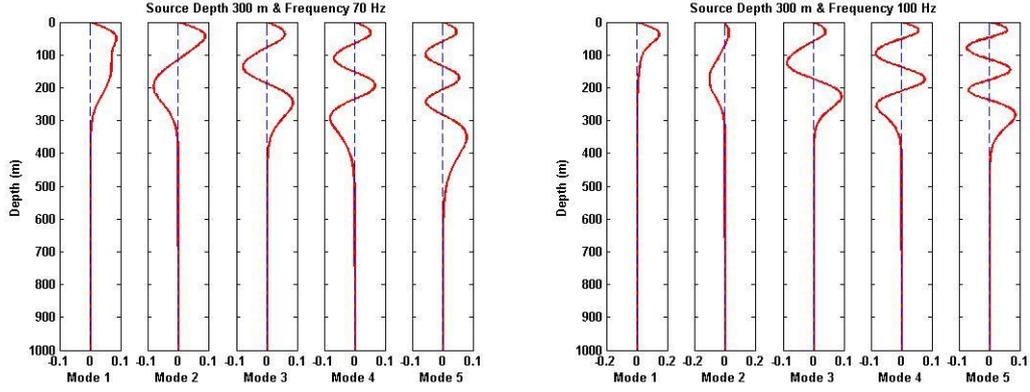

Figure 8 Vertical distribution of normal wave eigenfunctions with different frequencies and signals simulated under typical dual-channel SSP conditions in the Arctic

In summary, this article utilizes the vertical distribution of normal wave eigenfunctions under typical Arctic SSPs to explain the time-domain waveform and dispersion characteristics of simulated sound signals, including the low-frequency part of the normal wave reaching faster than the high-frequency part, and the existence of upper and lower frequency limits. It also explains the impact mechanism of changes in SSPs on both, including the cross phenomenon of dispersion structures caused by dual-channel SSPs.

## IV. INTRODUCTION TO THE EXPERIMENTAL PROCESS

Next, this article will provide a detailed introduction to an Arctic sound propagation experiment. From August 1, 2018 to August 25, 2018, a low-frequency long-range sound propagation experiment was conducted in the waters of the Chukchi Plateau and the Canadian Basin in the Arctic. Among them, the receiving equipment is a self-contained underwater acoustic signal acquisition system, and the sound source is a pulse sound source. During the experiment, seawater SSP measurements were carried out along the route. At the same time, satellite was used to obtain sea ice data of the experimental sea area during the experiment period. The underwater terrain data of the experimental sea area mainly relies on the ETOP database. Next, this paper will provide a detailed introduction in sequence.

### A. Acoustic signal receiving equipment

In this acoustic propagation experiment, the acoustic signal receiving equipment was deployed in the form of a submersible mooring buoy on the Chukchi Plateau. The submersible buoy system carried four self-contained underwater acoustic signal acquisition systems, numbered C144, F109, F180, and F181, all with a sampling rate of 16 k. At the same time, thermometers are installed near the sound receiving equipments to record the working depth. Table 1 shows the depth and signal acquisition methods of the four receiving devices. Due to the non-continuous operation mode of the signal acquisition methods of the second, third, and fourth devices, the devices are in a dormant state when the sound signal arrives, and no sound propagation signal is collected. Therefore, this article will only analyze the data of the first receiving device. The working depth of the first set of equipment is 372m, and according to Figure 3, it can be seen that this set of equipment is located in the surface seawater waveguide of the deep Arctic sea.

Table 1 The depth and data acquisition method of acoustic signal receiving equipment

| Equipment number | Depth | Data acquisition method |
| --- | --- | --- |
| C144 | 372m | Continuous acquisition |
| F109 | 373m | Discontinuous acquisition |
| F180 | 1170m | Discontinuous acquisition |
| F181 | 1789m | Discontinuous acquisition |

## B. Sound source equipment

In this sound propagation experiment, the sound source used was an impulsive sound source with a nominal depth of 300m. The impulsive sound source was dropped during the navigation of the experimental ship, and the location and time information of the sound source were shown in Table 2. From the table, it can be seen that five remote sound propagation experiments were conducted in this experiment, including the Chukchi Plateau and the Canadian Basin, with propagation distances of approximately 105km, 199km, 864km, 1066km, and 1093km, respectively. The maximum propagation distance exceeded 1000km. In addition, the latitude of NS station, S11 station, and S22 station is concentrated around 75 °N, and the longitude changes from 160 °W to 166 °W. Therefore, these two sound propagation experiments can be approximated as sound propagation experiments along 75 °N from west to east. The latitude from NS station to ICE05, ICE06, and ICE08 stations changes from 75 °N to 85 °N, while the longitude changes from 160 °W to 167 °W. Due to the greater distance variation corresponding to latitude changes in high latitude areas compared to longitude changes, these three sound propagation experiments can be similar to sound propagation experiments along 160 °W and from north to south, that is, ultra long range sound propagation experiments from the central ice zone at 85 °N to the Chukchi Plateau at 75 °N. During the propagation process, the seawater SSP changes from the typical SSP of the central ice zone to the dual-channel SSP of the Chukchi Plateau.

Table 2 Location and time information of sound signal receiving equipment and sound source

|  | Deployment time | Station name | Propagation range | Latitude | Longitude |
| --- | --- | --- | --- | --- | --- |
| Receiving equipment | 2018/08/01 | NS | 0km | 75.0 °N | 159.8 °W |
| Source 1 | 2018/08/02 | S11 | 105km | 74.7 °N | 163.3 °W |
| Source 2 | 2018/08/03 | S22 | 199km | 74.6 °N | 166.6 °W |
| Source 3 | 2018/08/15 | ICE05 | 864km | 82.6 °N | 167.3 °W |
| Source 4 | 2018/08/23 | ICE08 | 1066km | 84.6 °N | 161.9 °W |
| Source 5 | 2018/08/20 | ICE06 | 1093km | 84.8 °N | 165.9 °W |

During the experiment, raw sound signals were collected at close range from the sound source signals of ICE05, ICE08, and ICE06 stations, as shown in Figure 9.

From the figure, it can be seen that the impulsive sound source signal used in this experiment has a relatively stable waveform structure, with a high source level in the frequency range of 20Hz to 100Hz.

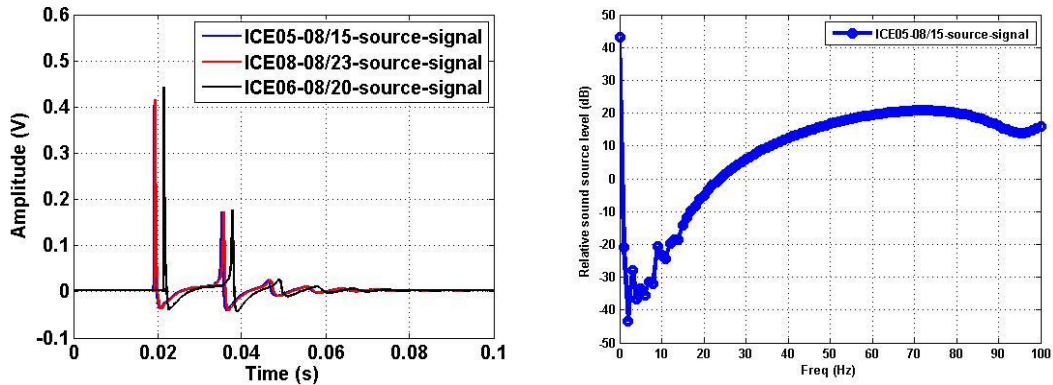

Figure 9 The original signal and spectrum analysis of impulsive sound sources

## C. Seafloor topography

Figure 10 shows the seabed topography of the experimental sea area, as well as the location of the sound sources, reception location, and sound propagation paths. From the figure, it can be seen that the sound propagation path from S11 and S22 stations to NS station changes from shallow sea with a depth of several hundred meters to deep sea with a depth of about 2000 meters. The sound propagation path from ICE05, ICE06, and ICE08 stations to NS station changes from deep sea with a depth of about 4000 meters to deep sea with a depth of about 2000 meters, but the propagation path is affected by the Chukchi Plateau seamounts in the middle. Therefore, this experiment can deeply analyze the impact of seabed terrain on sound propagation signals through the sound propagation paths under these complex seabed terrains.

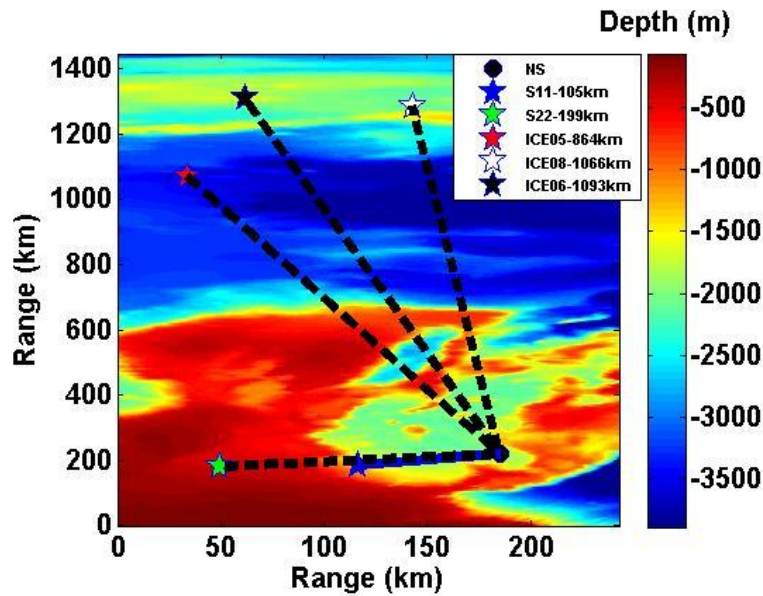

Figure 10 The seabed topography based on ETOP database, sound source location, reception location, and sound propagation path

## D. Sea ice coverage

Figure 11 shows the temporal variation of sea ice density observed by satellites in the experimental sea area during the experimental period. It can be seen from the figure that the sound propagation path from S11 and S22 stations to NS station is basically not covered by sea ice, while the sound propagation path from ICE05, ICE06, and ICE08 stations to NS station is mostly covered by sea ice, Therefore, this experiment can compare and analyze the impact of sea ice cover on the characteristics of sound signals through several different sound propagation paths under different sea ice cover conditions.

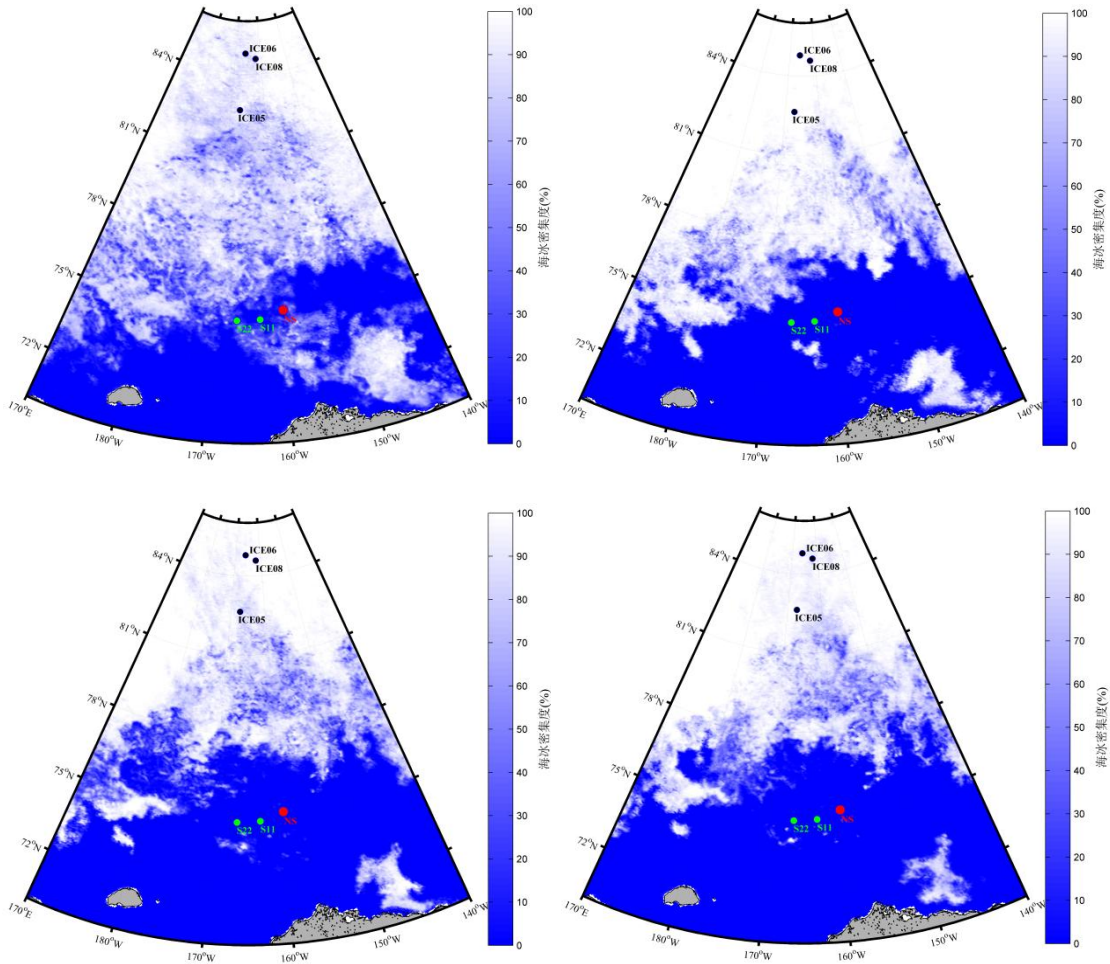

Figure 11 The images of sea ice density in the experimental sea area based on satellite observations during the experimental period, sound source locations and reception location

## E. Sea water sound speed profile

During this experiment, the SSPs of 29 stations were measured along the navigation route of the experimental vessel. The relative positions of the SSP measurement stations, the sound source stations, and the receiving stations are shown in Figure 12. Although the SSP measurement was not carried out along the sound propagation path in this experiment, the location of the sampling station in the experiment is conducive to analyzing the trend of the SSP changes with longitude and latitude in the sea area. Figure 13 shows the spatial variation pattern of SSPs in four segments of the experimental ship's navigation route. As shown in Figure 13, in the direction of longitude, the SSP changes from a dual-channel SSP in the Canadian Basin to a single channel positive gradient SSP in the East Siberian Sea from west to east; In the latitude direction, the SSP changes from a dual-channel SSP in the Canadian Basin to a single channel positive gradient SSP in the central ice region from south to north. Therefore, through this experiment, the impact of complex SSP changes on the characteristics of sound signals along the sound propagation path can be analyzed.

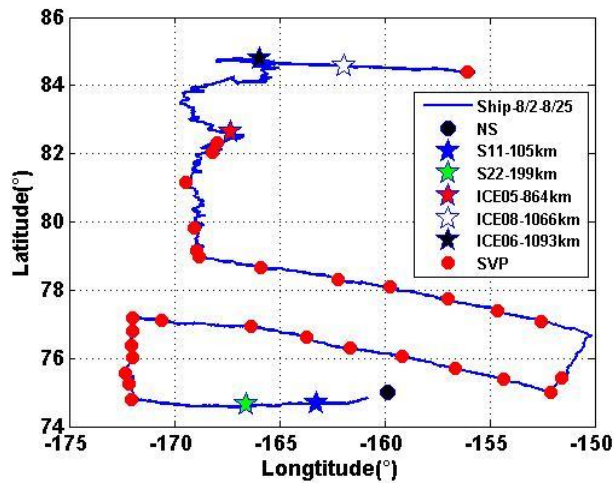

Figure 12 The locations of SSP measurement stations, sound source locations, and reception location

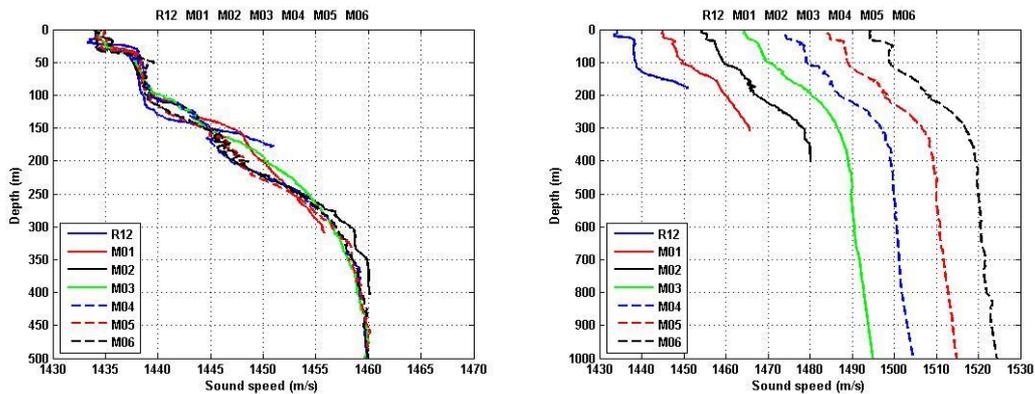

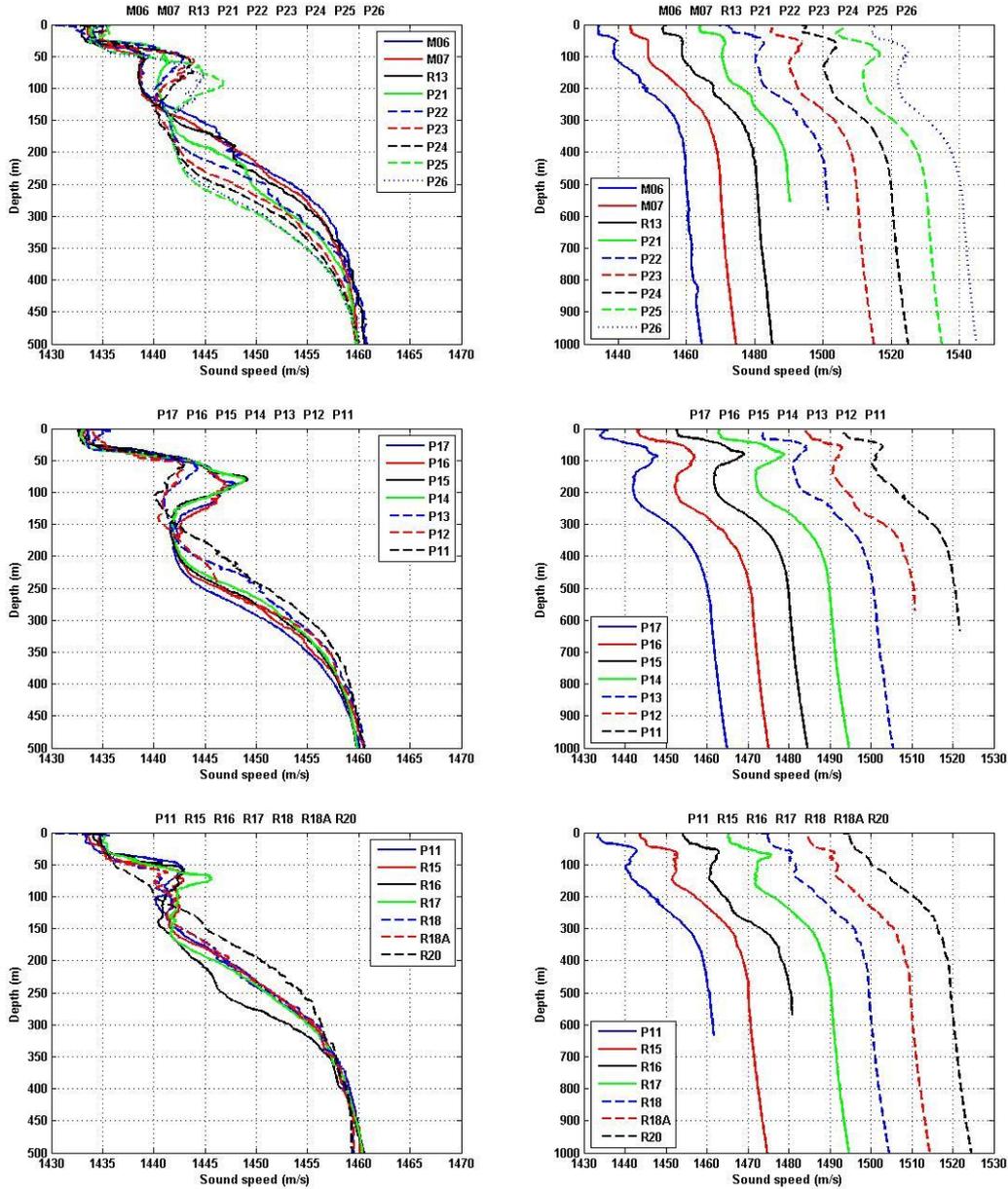

Figure 13 The variation of SSP with space on the four sections of the experiment ship's route

## V. EXPERIMENTAL DATA PROCESSING

Next, this article will conduct a detailed analysis of the received acoustic signal in the experiment, including time-domain waveform and time-frequency analysis of the acoustic signal, as well as using warping changes to separate normal modes to extract dispersion curves. At the same time, simulation of received acoustic signals is carried out based on the actual acoustic propagation environment, and compared with experimental signals. The purpose of conducting these analysis works is to verify the underwater sound propagation laws in the Arctic and the sound field effects of the Arctic marine environment, and to verify the effectiveness of using warping transform to achieve refractive normal mode separation in the Arctic.

## A. Pulse sound signal with a distance of 105 km from S11 station

Figure 14 shows the seabed terrain of the sound propagation path from S11 station to NS station, based on the ETOP seabed terrain database. It can be seen that the seabed depth changes from about 1300 m to about 1900 m within a distance of 105 km, and there is a seabed mountain with a depth of about 1200m at about 75km. From the figure, it can be seen that the distance between these two stations is relatively short, and the fluctuation of the seabed terrain is relatively small. So the corresponding horizontal variation of the SSP should also be relatively small. In addition, due to the ETOP database not being measured data, there may be deviations in the seabed topography of the sound propagation path.

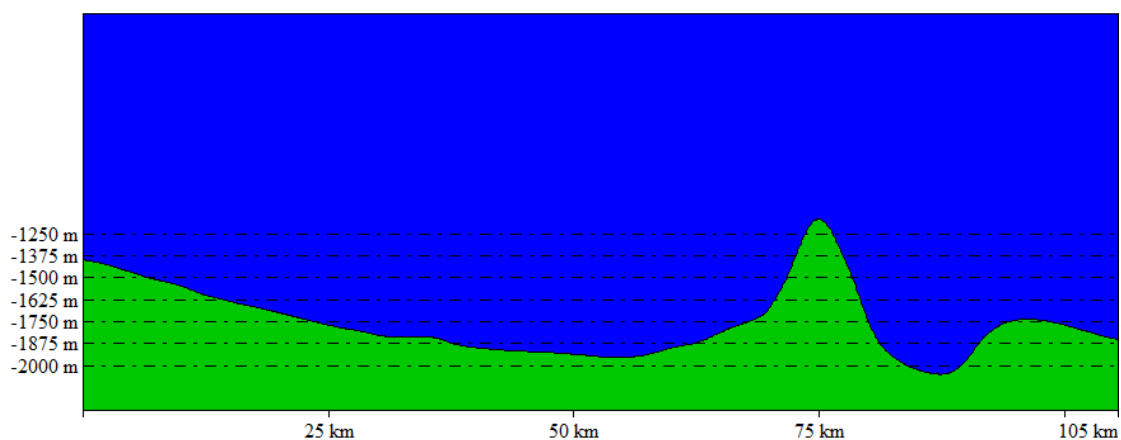

Figure 14 The terrain of the seabed from S11 station to NS station based on ETOP database

Figure 15 shows the time-domain waveform and time-frequency analysis of the received acoustic signal. From the figure, it can be seen that the main part of the acoustic signal lasts for about 1 second, followed by some weak signals, which should be formed by sound rays with multiple sea bottom reflections. From the time-frequency analysis graph, it can be seen that the acoustic signal mainly consists of two parts, including multiple broadband multipath pulse signals without dispersion and about eleven normal modes with dispersion. Among them, broadband multipath arrival signals are formed by multiple sound rays that reverse through deep water, and due to the fast sound speed of the water they pass through, they arrive earliest. The dispersion structure of normal mode has the characteristic of increasing frequency with time, which is consistent with the simulation results. Warping transformation was carried out on the received sound signal. From the figure, it can be seen that each normal mode has become a single frequency signal. Through band-pass filtering and inverse warping transformation, six strong normal modes and their corresponding dispersion curves were extracted. The extracted dispersion curves are completely consistent with the original signal, verifying the effectiveness of refractive normal mode warping transformation and its application in extracting dispersion curves. In addition, from time-frequency analysis, it can be seen that each normal mode has a clear upper frequency limit.

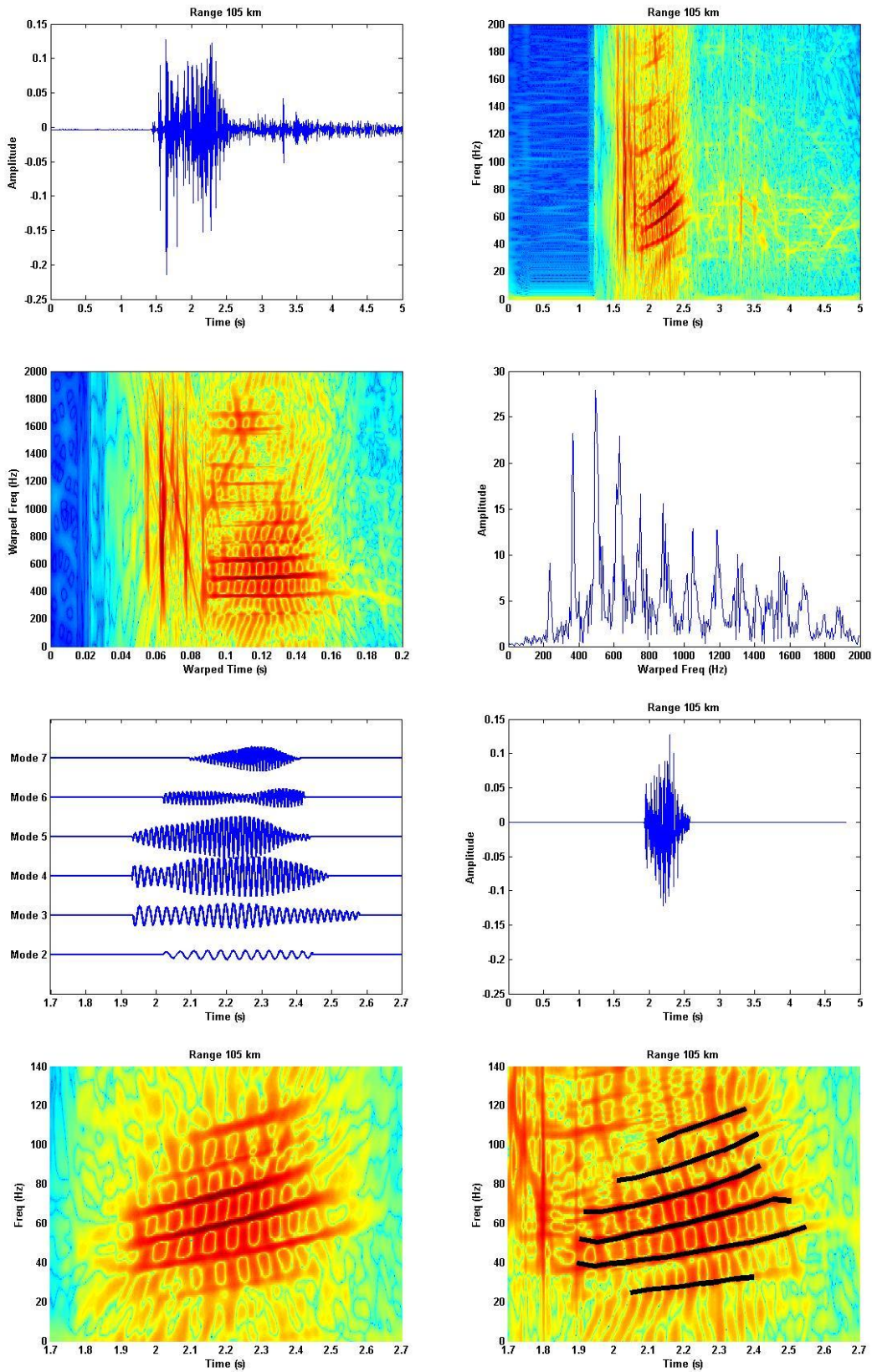

Figure 15 Time domain waveform and time-frequency analysis of acoustic signals with a propagation distance of 105km from S11 station to NS station, as well as the warping transformation, normal mode separation, and dispersion curve extraction process for the acoustic signal

Based on the seabed terrain along the sound propagation path shown in Figure 14, a two-dimensional sound transmission loss calculation is performed using the parabolic equation sound field model Ram, assuming a sound source frequency of 50 Hz. Figure 15 shows the calculation results, and it can be seen from the figure that the two-dimensional sound field is mainly composed of a surface normal mode sound field and a deep-reversed multi-path sound rays sound field, which corresponds to the main characteristics of the experimental sound signal.

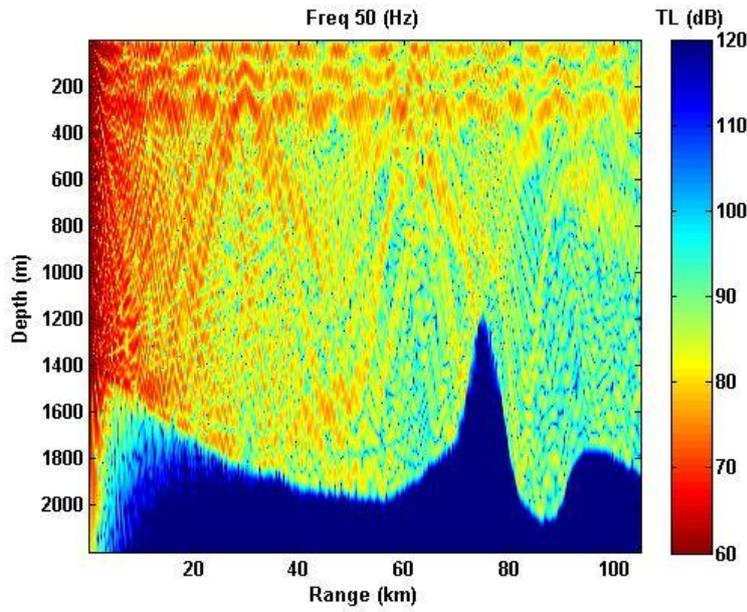

Figure 16 The calculation result of two-dimensional sound transmission loss from S11 station to NS station using the parabolic equation sound field model Ram and the seabed topography from ETOP database

Next, this paper uses the parabolic equation model Ram to conduct time-domain waveform simulation of received acoustic signals. Due to the lack of measurement of the SSP on the path from the sound source to the receiver during the experiment, referring to Figure 12, the SSPs of nearby stations P21 and P22 were used instead. Figure 17 shows the time-domain waveform and time-frequency analysis of the simulated acoustic signal, as well as the comparison of the dispersion curve extracted from the experiment with the simulated acoustic signal. From the figure, it can be seen that the simulated acoustic signals under these two SSPs are basically consistent, and the dispersion structure of the simulated acoustic signals is basically consistent with the experimental results. The simulated acoustic signal and experimental signal have the same upper limit of normal mode frequency, but there is a certain difference in the lower limit of frequency between the two, which may be caused by inaccurate seabed topography.

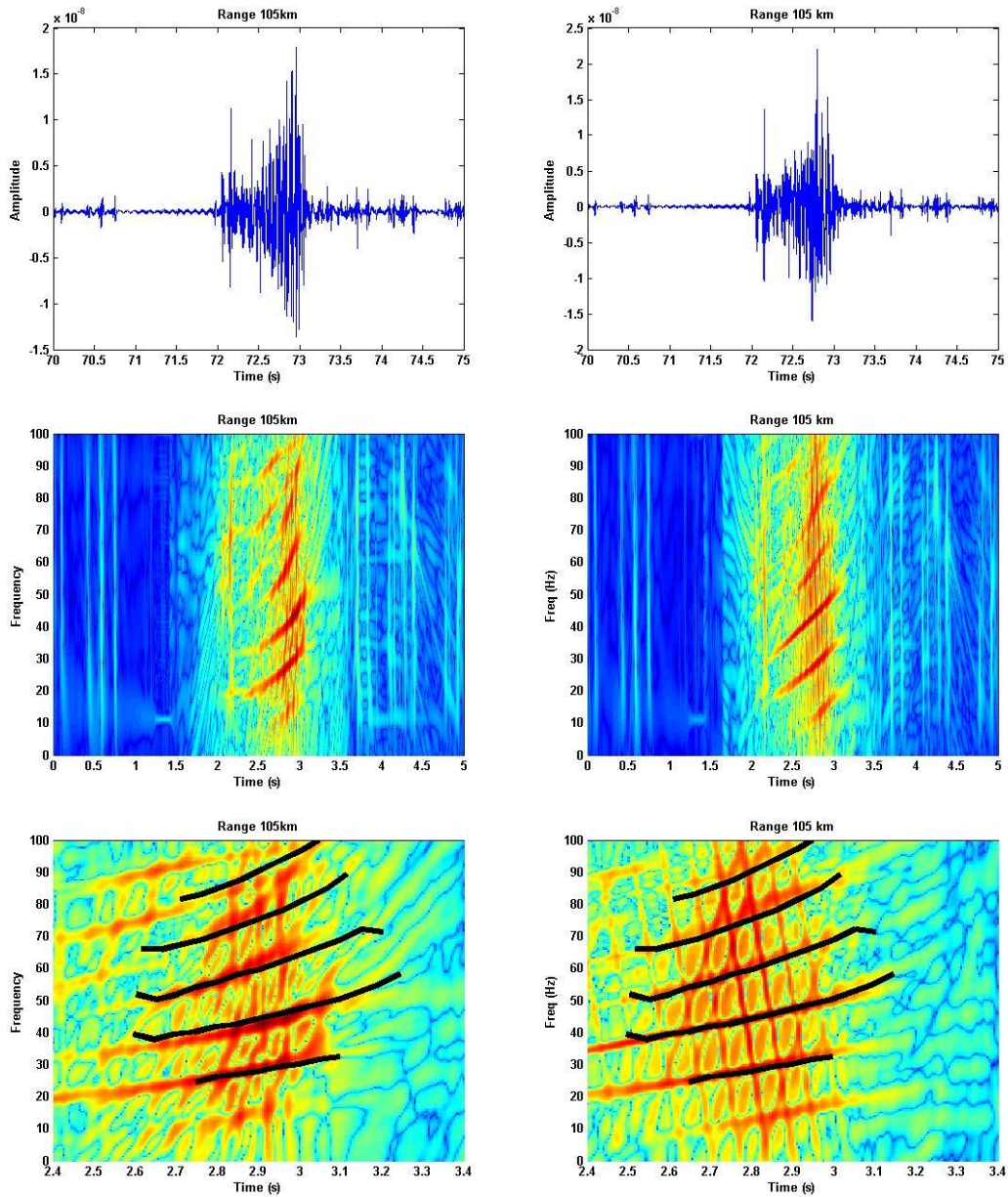

Figure 17 The time-domain waveform and time-frequency analysis of simulated acoustic signals from S11 station to NS station, as well as comparison with frequency dispersion curves extracted from measured signals

By comparing the SSPs of stations P21 and P22, as shown in Figure 18, it can be seen that there are certain differences between the two, especially in the intensity of the dual channel. In order to make the simulation environment more consistent with the experimental environment, both P21 and P22 SSPs were used in the simulation to calculate the sound signal when the SSP changes with horizontal distance. The simulation results are shown in Figure 19. From the figure, it can be seen that the time-domain waveform of the simulated acoustic signal is basically consistent with the measured acoustic signal, and the dispersion structure in time-frequency analysis is also basically consistent with the dispersion curve of the measured signal. However, there is a certain difference between the two in the lower frequency limit of normal mode. The simulated acoustic signal has a lower frequency limit, and the actual seabed terrain may be shallower than the simulated seabed terrain.

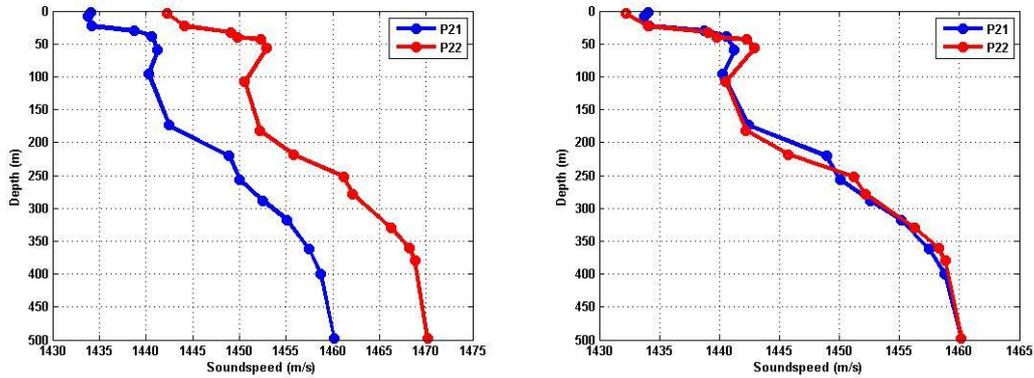

Figure 18 The SSPs of two stations P21 and P22 used in the simulation and their comparison

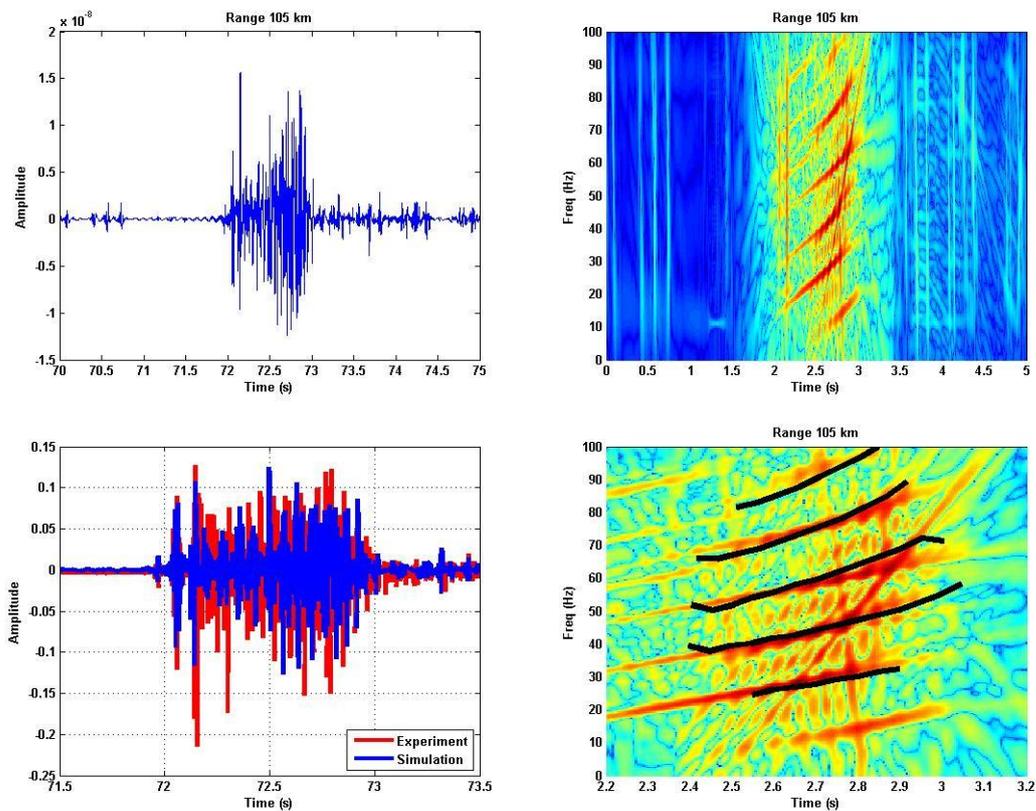

Figure 19 The simulation result of the signal time-domain waveform and time-frequency analysis based on horizontal variation of SSP conditions, and comparison with experimental dispersion curves

From this example of sound propagation, it can be seen that under the typical deep sea SSP conditions in the Arctic, the received sound signal includes multipath arrival and multiple normal modes arrival, and there are also corresponding large-depth reversed sound rays and surface normal modes in the two-dimensional sound field. Among them, the dispersion curve of normal modes has the characteristic of increasing frequency with time. From the experimental data and processing results, it can be seen that using the warping transformation of refractive normal modes can achieve the separation of normal modes and the extraction of dispersion curves. In the absence of sea ice cover, the upper limit of normal mode frequency for simulated acoustic signals and experimental signals is consistent, but inaccurate seabed topography can affect the lower limit of frequency.

## B. Pulse sound signal with a distance of 199 km from S22 station

Figure 20 shows the changes in the seabed terrain from S22 station to NS station. It can be seen from the figure that the seabed terrain can be roughly divided into three sections within the range of 200km. The sea depth of the first section is maintained at around 400m within the range of 0km to 50km, the sea depth of the second section changes from about 400m to about 2000m within the range of 50km to 150km, and the sea depth of the third section is maintained at around 2000m within the range of 150km to 200km. Therefore, this sound propagation path is a transitional sea area sound propagation from shallow to deep sea under the conditions of the Arctic SSP, which is a good example for studying the influence of seabed topography on sound propagation.

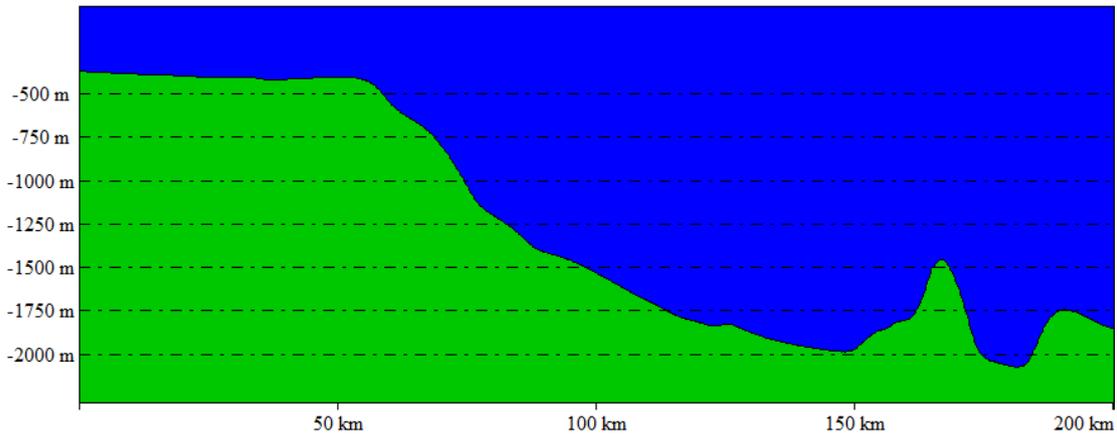

Figure 20 The terrain of the seabed from S22 station to NS station based on ETOP database

Figure 21 shows the time-domain waveform and time-frequency analysis of the received acoustic signal. From the figure, it can be seen that the duration of the acoustic signal is about 1.5 seconds, followed by some weak signals, which should be formed by sound rays with multiple sea bottom reflections. From the time-frequency analysis graph, it can be seen that the acoustic signal is mainly composed of multiple normal modes with dispersion characteristics, and the dispersion structure of the normal mode has the characteristic of increasing frequency with time. Compared with the time-frequency analysis of the acoustic signal in figure 15, the signal in figure 21 does not have broadband multipath pulse signals without dispersion characteristics. This is due to the presence of shallow sea terrain in figure 20, which prevents the arrival of multipath large-depth reversed acoustic rays on this propagation path. Meanwhile, due to the presence of shallow sea bottom topography, the acoustic signal in figure 21 not only has a clear upper limit of normal mode frequency, but also has a clear lower limit of normal mode frequency. This is because the eigenfunctions of low-frequency part of normal modes are obscured by shallow sea terrain. Warping transformation was carried out on the received sound signal. From the figure, it can be seen that each normal mode has become a single frequency signal. Through bandpass filtering and inverse warping transformation, strong second to sixth normal modes and corresponding dispersion curves were extracted, which are completely consistent with the original signal.

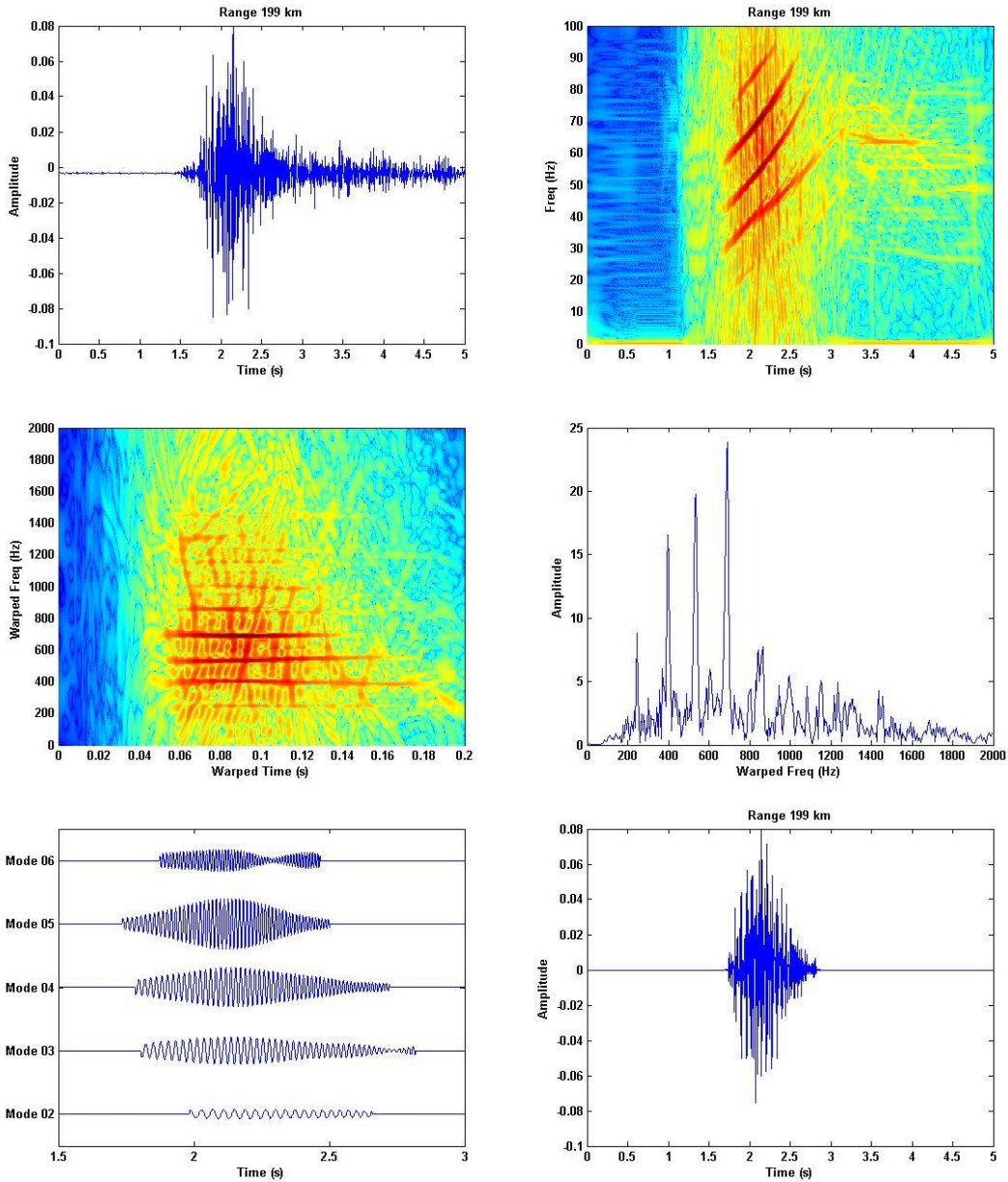

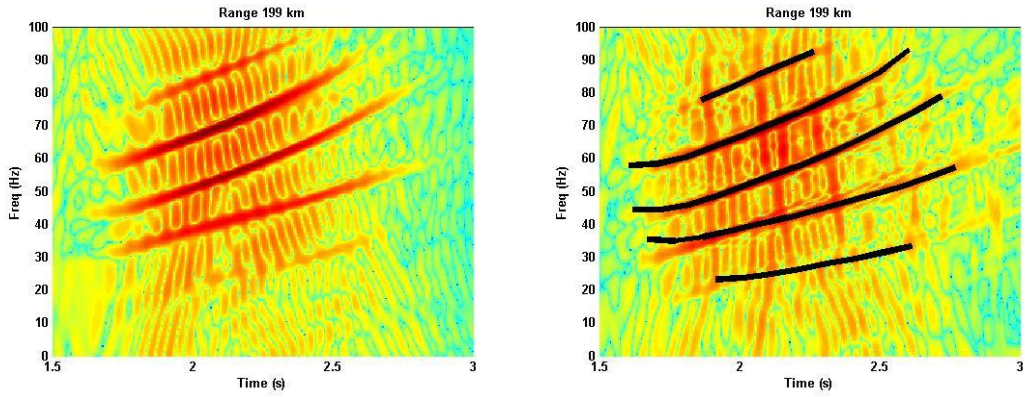

Figure 21 Time domain waveform and time-frequency analysis of acoustic signals with a propagation distance of 199 km from S22 station to NS station, as well as the warping transformation, normal mode separation, and dispersion curve extraction process for the acoustic signal

Based on the seabed terrain along the sound propagation path shown in Figure 20, a two-dimensional sound transmission loss calculation is performed using the parabolic equation sound field model Ram, assuming a sound source frequency of 50 Hz. Figure 22 shows the calculation results. From the figure, it can be seen that the sound field energy propagates for 50km in shallow waters and begins to reach deep waters, after that, the sound field energy remains concentrated within the 400m waveguide on the surface, with weak energy reaching at large depths. This corresponds to the phenomenon in the experimental signal where only normal mode waves arrive and no large-depth reversed sound rays arrive. Therefore, under typical SSP conditions in the Arctic, there is no common downhill debris flow effect in the transition zone from shallow to deep sea, and the sound field energy is still concentrated on the surface of the seawater. The shallow sea floor also blocks the low-frequency parts of normal modes.

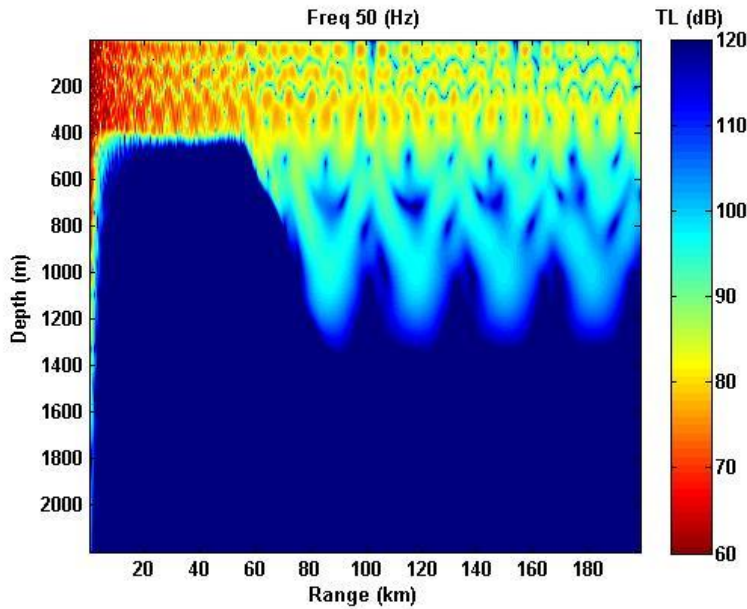

Figure 22 The calculation result of two-dimensional sound transmission loss from S22 station

to NS station using the parabolic equation sound field model Ram

Next, this paper uses the parabolic equation model Ram to conduct time-domain waveform simulation of received acoustic signals. Due to the lack of measurement of the SSP from the source to the received propagation path during the experiment, referring to figure 12, the SSPs of stations P23 and P26 with typical deep-sea dual-channel characteristics were used instead. Figure 23 shows the time-domain waveform and time-frequency analysis of the simulated acoustic signal, as well as the comparison of the dispersion curve extracted from the experiment with the simulated acoustic signal. From the figure, it can be seen that the time-domain waveform of the simulated acoustic signal under these two SSPs is relatively consistent with the experimental signal, but the dispersion structure of the simulated acoustic signal differs significantly from the dispersion curve extracted from the experiment.

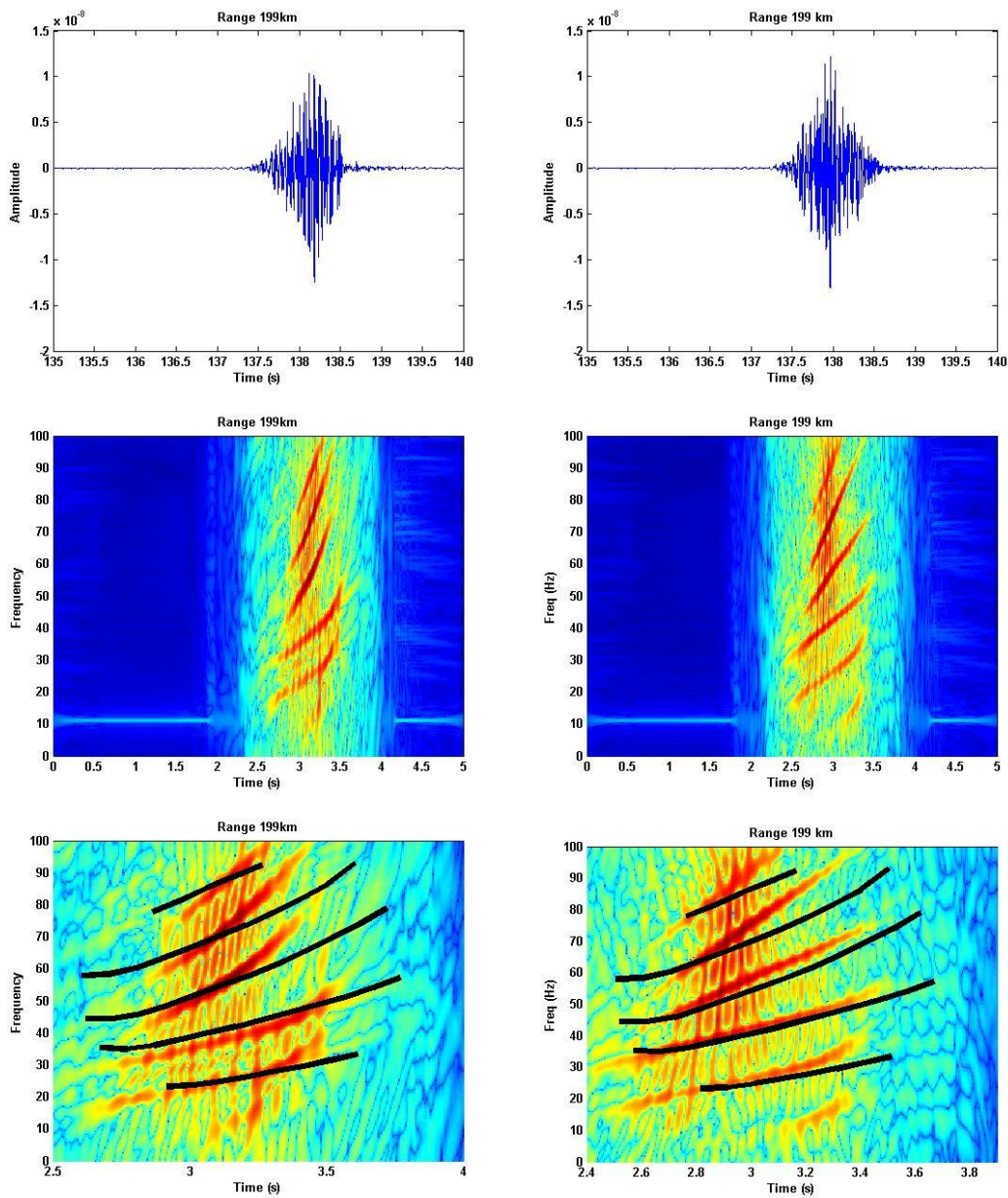

Figure 23 The time-domain waveform and time-frequency analysis of simulated acoustic

signals from S22 station to NS station, as well as comparison with frequency dispersion curves extracted from measured signals

By comparing the SSPs of stations R13, P21, and P22, as shown in figure 24, it can be seen that there are certain differences among the three, including the intensity and width of the dual channel. In order to make the simulation environment more consistent with the experimental environment, three SSPs, R13, P21, and P22, were simultaneously used in the simulation to calculate the sound signal when the SSP changes with distance. The simulation results are shown in figure 25. From the figure, it can be seen that the time-domain waveform of the simulated acoustic signal is basically consistent with the measured acoustic signal, and the dispersion structure in time-frequency analysis is also basically consistent with the dispersion curve of the measured signal. At the same time, the simulated acoustic signal and experimental signal have a consistent upper limit of normal mode frequency, because this acoustic propagation path is not covered by sea ice on the surface. In addition, the lower frequency limits of the two are also basically the same, which should be a result of more accurate seabed topography.

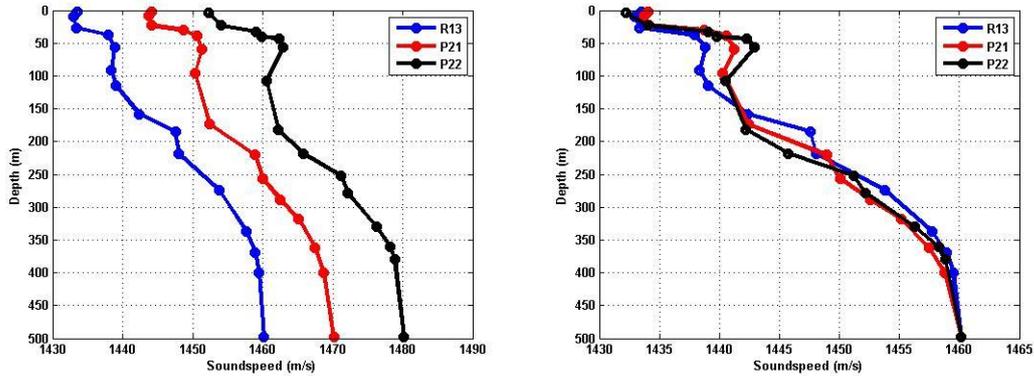

Figure 24 The SSPs of R13, P21 and P22 used in the simulation and their comparison

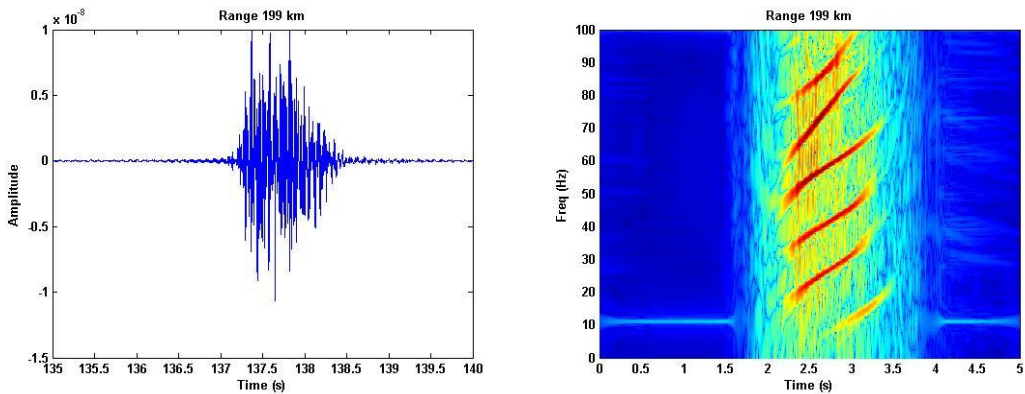

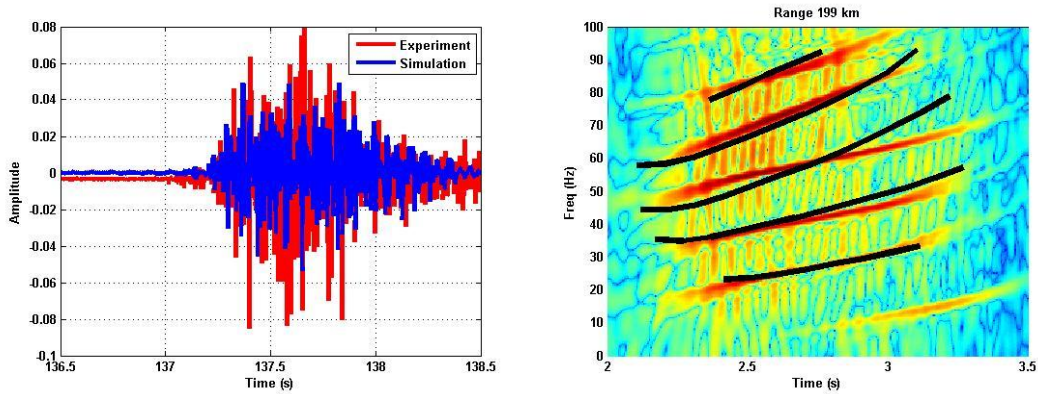

Figure 25 The simulation result of the signal time-domain waveform and time-frequency analysis based on horizontal variation of SSP conditions, and comparison with experimental dispersion curves

From this example of sound propagation, it can be seen that under typical deep sea SSP conditions in the Arctic, when sound signals are transmitted from shallow water to deep sea, the sound energy is mainly concentrated in the surface of the seawater, and there is almost no sound energy reaching large depths. At the same time, the sound signal mainly manifests as a structure where multiple normal modes arrive, and the dispersion structure of the normal mode has the characteristic of increasing frequency with time. In addition, due to the action of the shallow sea bottom, normal mode waves have a clear lower frequency limit.

## C. Pulse sound signal with a distance of 864 km from ICE05 station

Figure 26 shows the changes in the seabed terrain from ICE05 to NS station. It can be seen from the figure that the seabed terrain along the sound propagation path is relatively complex. The sound signal propagates from the Canadian Basin to the Chukchi Plateau, with a sea depth ranging from about 3500m to about 2000m, and there is also a seamount with a depth of about 1000m in the middle. At the same time, according to the SSPs measured in figures 12 and 13, it can be seen that the sound source is located in the central ice zone, and the reception is located in the Chukchi Plateau. The SSP on the sound propagation path changes from a single channel SSP in the deep sea to a dual-channel SSP in the deep sea, and the spatial variation of the SSP is relatively significant. In addition, according to the satellite images of the sea area at that time, as shown in figure 11, at least half of the sea area on the sound propagation survey line is covered by sea ice. In summary, the sound propagation conditions of this example are relatively complex, involving the horizontal changes in seabed topography, SSP, and sea ice cover.

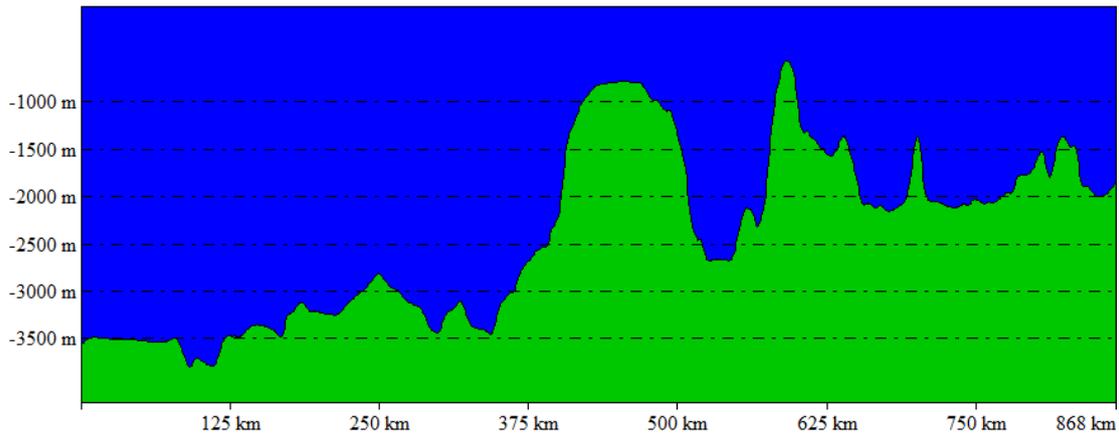

Figure 26 The terrain of the seabed from ICE05 station to NS station based on ETOP database

Figure 27 shows the time-domain waveform and time-frequency analysis of the received acoustic signal. From the figure, it can be seen that the duration of the acoustic signal exceeds 6 seconds, mainly composed of the arrival of multiple normal modes with dispersion characteristics. There are obvious first to fifth normal modes in the acoustic signal, and the dispersion structure of the mode wave still maintains the characteristic of increasing frequency with time. Among them, the energy of the first normal mode is very weak, possibly due to the obstruction of shallow seamounts on the propagation path. At the same time, each normal mode wave has a clear upper frequency limit, which is because at least half of the sea surface on the sound propagation path is covered by sea ice, causing the high-frequency part of the normal mode to be attenuated by the ice-water interface. Conducting warping transformation on the received sound signal, it can be seen from the figure that although the transformed sound signal has not become a theoretical single frequency signal, the second to fifth normal modes can still be separated using a bandpass filter and the corresponding dispersion curve can be extracted. The extracted dispersion curve is completely consistent with the time-frequency analysis of the measured signal, while the first normal mode cannot be separated due to its weak energy.

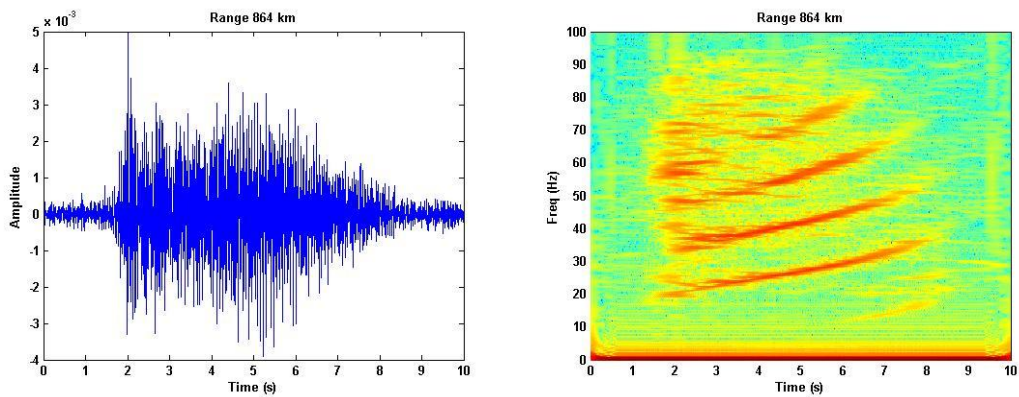

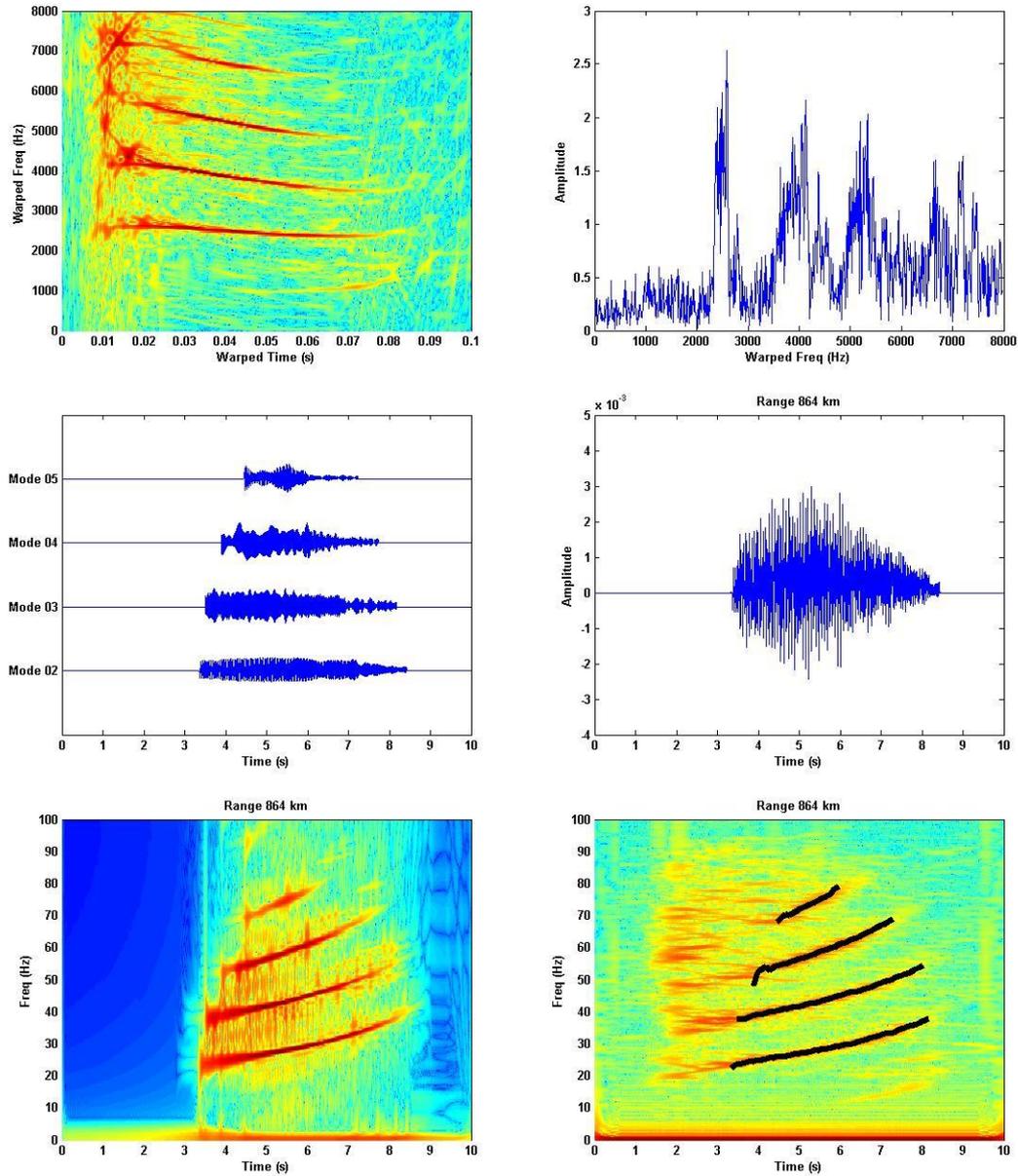

Figure 27 Time domain waveform and time-frequency analysis of acoustic signals with a propagation distance of 864 km from ICE05 station to NS station, as well as the warping transformation, normal mode separation, and dispersion curve extraction process for the acoustic signal

Based on the seabed terrain along the sound propagation path shown in Figure 26, a two-dimensional sound transmission loss calculation is performed using the parabolic equation sound field model Ram, assuming a sound source frequency of 50Hz. Figure 28 shows the calculation results. From the figure, it can be seen that the complex seabed terrain mainly obstructs the propagation of large-depth reversed sound rays, and has a small impact on the normal mode waves on the surface. Therefore, the main component of the received sound signal at long propagation ranges is the normal modes, which is consistent with the experimental results of the received sound signal. Therefore, under the typical SSP conditions in the Arctic, the complex and variable ocean terrain makes the sound signals mainly consist of multiple low-frequency normal modes in the surface water after long-distance propagation.

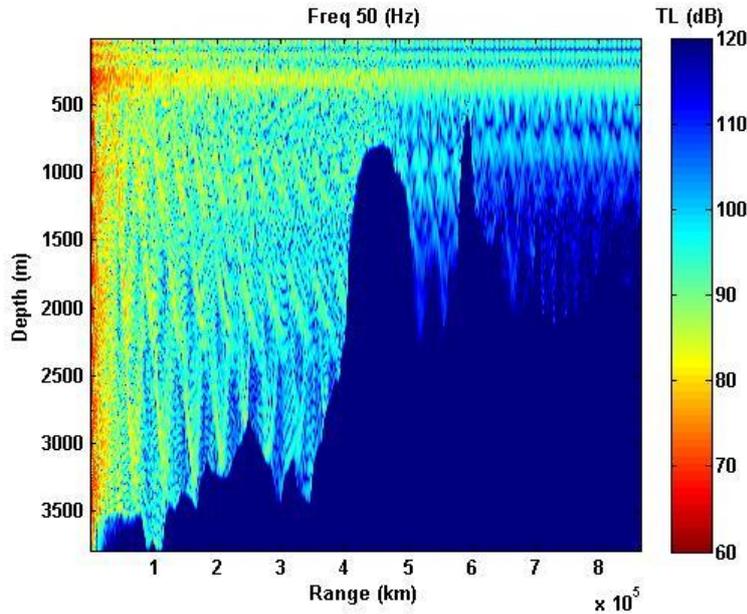

Figure 28 The calculation result of two-dimensional sound transmission loss from ICE05 station to NS station using the parabolic equation sound field model Ram

For the sound propagation problem from ICE05 station to NS station, due to the propagation distance of the sound signal being too far, the sound source station is located in the central ice zone at latitude of about 82.5 °N, and the receiving station is located in the Chukchi Plateau at latitude of about 75 °N, the SSPs of stations near the propagation path can be obtained using Figures 12 and 13, as shown in Figure 29. From the figure, it can be seen that the SSP undergoes significant changes on the propagation path, from a single channel SSP to a dual-channel SSP.

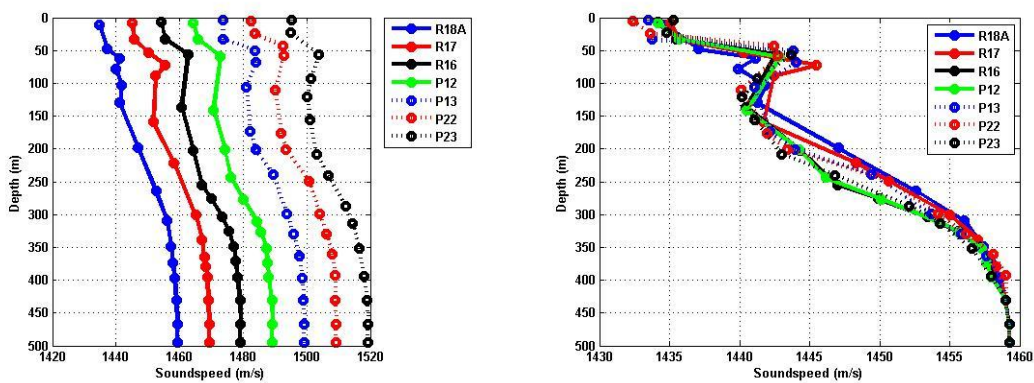

Figure 29 SSPs of seven nearby stations on the sound propagation path from ICE05 sound source station to NS receiving station

In order to analyze the impact of changes in SSPs on long-distance propagation of sound signals, this article uses two types of SSPs for simulation. The first is to assume that the SSP remains unchanged horizontally and is fixed as a single-channel SSP in the central ice region of the Arctic where the sound source is located. The second is to assume that the SSP changes horizontally and interpolate the seven SSPs in figure 29 to obtain a sound velocity field that varies with distance. Figure 30 shows the time-domain waveform and time-frequency analysis of the simulated acoustic signal in two different scenarios, as well as a comparison with the time-domain waveform of the experimental signal and the extracted dispersion curve. The differences between the simulated acoustic signals in the two scenarios are significant. From the figure, it can be seen that the time-domain waveform and dispersion structure of the acoustic signal calculated considering the horizontal variation of the SSP are more consistent with the experimental results, indicating that the influence of the horizontal variation of the SSP needs to be considered in the study of long-distance sound propagation in the Arctic Ocean. The upper limit of the normal mode frequency of the simulated acoustic signal is significantly greater than the upper limit of the normal mode frequency of the experimental acoustic signal, and this is because the effect of sea ice cover is not considered in the acoustic signal simulation, resulting in the high-frequency part of the simulated acoustic signal not being attenuated, therefore resulting in inconsistency between the two. In addition, the simulated acoustic signal contains the first mode with strong energy, which is also inconsistent with the experimental acoustic signal, and this is because accurate seabed topography and acoustic parameters were not measured in this experiment. In summary, in the long-distance sound propagation process under the Arctic Ocean ice, changes in SSPs, seabed topography, and sea ice cover all have a significant impact on sound propagation, but the ways of action are different.

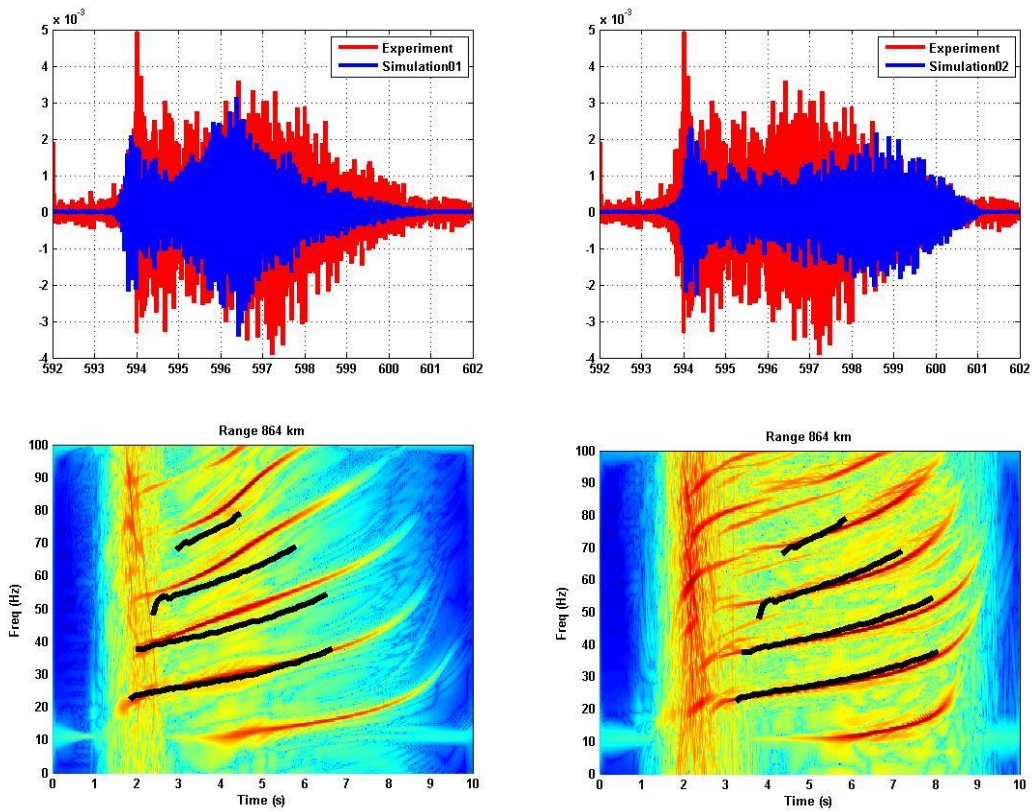

Figure 30 The time-domain waveform and time-frequency analysis of simulated acoustic

## D. Pulse sound signal with a distance of 1066 km from ICE08 station

Figure 31 shows the changes in seabed terrain from ICE08 to NS station. It can be seen from the figure that the seabed terrain along the sound propagation path is relatively complex. The sound signal starts from the Mendeleev Ridge, passes through the Canadian Basin, and finally reaches Chukchi Plateau. Among them, the Mendeleev Ridge section has a sea depth of about 2000m, the Canadian Basin section has a sea depth of about 3500m, and the Chukchi Plateau section has a sea depth of about 2000m. Meanwhile, according to the satellite images of the sea area at that time, as shown in figure 11, more than two-thirds of the sea area on the sound propagation line is covered by sea ice, so it is necessary to consider the impact of sea ice. In addition, according to the SSPs measured in figures 12 and 13, it can be seen that the sound source is located in the central ice zone, and the reception is located in the Chukchi Plateau. The SSP on the sound propagation path changes from a deep-sea single channel to a deep-sea dual channel, and the spatial variation of the SSP is relatively significant. In summary, the sound propagation conditions of this example are relatively complex, involving the horizontal changes in seabed topography, sound velocity profile, and sea ice cover.

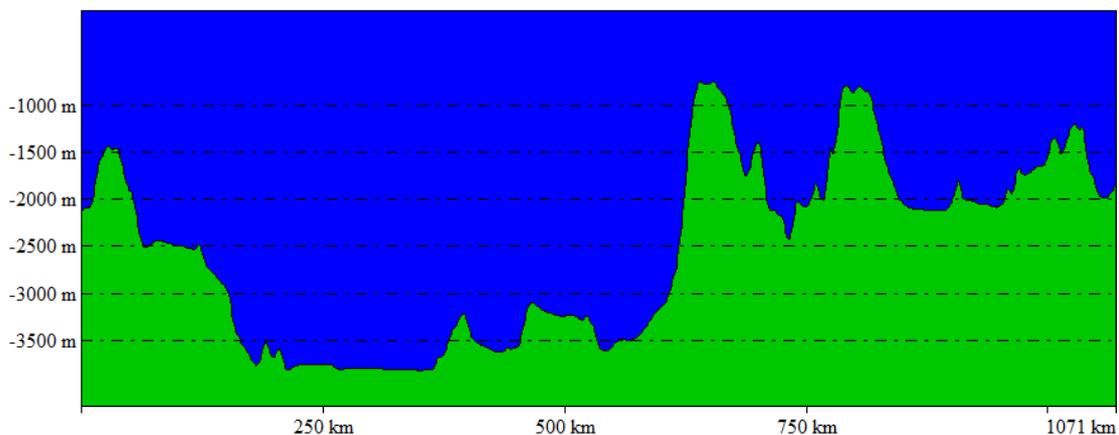

Figure 31 The terrain of the seabed from ICE08 station to NS station based on ETOP database

Figure 32 shows the time-domain waveform and time-frequency analysis of the received acoustic signal. From the figure, it can be seen that the duration of the acoustic signal exceeds 8 seconds, mainly composed of the arrival of multiple normal modes with dispersion characteristics. Compared with figure 27, the duration of the acoustic signal with a propagation distance of 1066km is significantly longer than that of the acoustic signal with a propagation distance of 864km, which is consistent with the results in the simulation section. There are obvious first to fifth mode modes in the acoustic signal, and the energy of the first normal mode is relatively stronger than that of the received acoustic signal with a propagation distance of 864km, and this may be due to the relatively weak shielding effect of the seamount from ICE08 station to NS station on the acoustic signal. At the same time, each normal mode wave has a clear upper frequency limit, which is because more than 2/3 of the sea surface along the sound propagation path is covered by sea ice, resulting in the high-frequency part of the normal mode wave being attenuated by the ice water interface. Then warping transformation is carried out on the received sound signal. From the figure, it can be seen that although the transformed sound signal has not become a theoretical single frequency signal, the first five normal modes can still be separated using a bandpass filter and the corresponding dispersion curve can be extracted. The extracted dispersion curve is completely consistent with the time-frequency analysis of the actual measured signal.

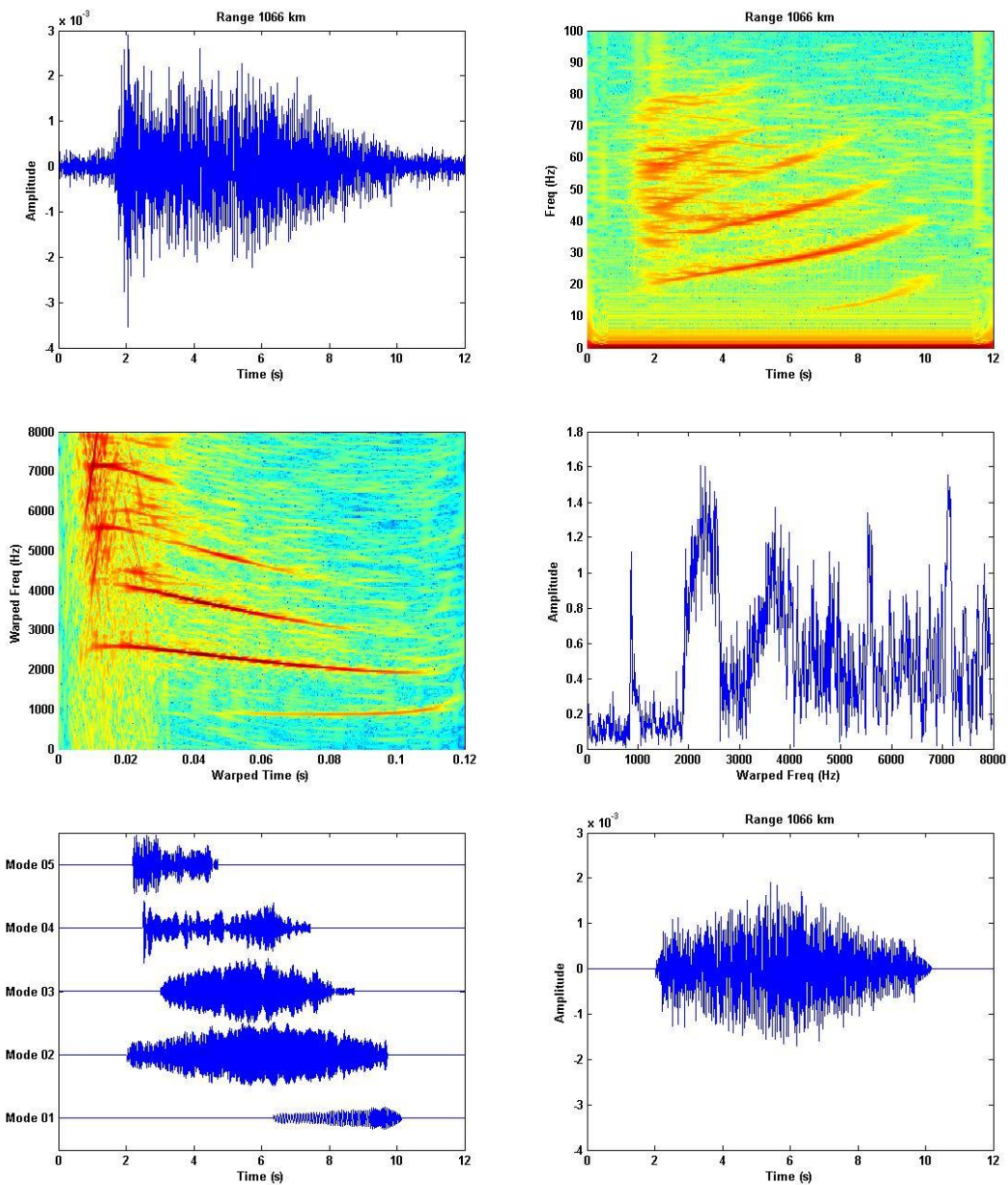

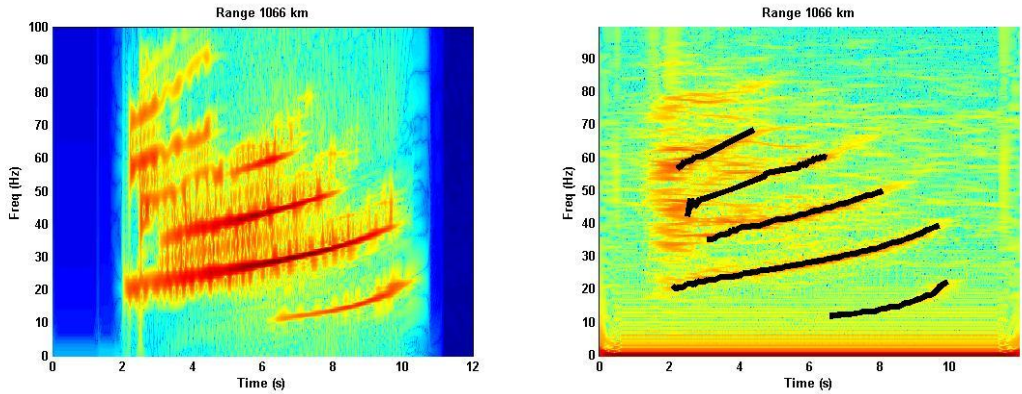

Figure 32 Time domain waveform and time-frequency analysis of acoustic signals with a propagation distance of 1066 km from ICE08 station to NS station, as well as the warping transformation, normal mode separation, and dispersion curve extraction process for the acoustic signal

Based on the seabed terrain along the sound propagation path in figure 31, a two-dimensional sound transmission loss calculation is performed using the parabolic equation sound field model Ram, assuming a sound source frequency of 50Hz. Figure 33 shows the calculation results, which are similar to those shown in figure 28. From the figure, it can be seen that the complex seabed terrain mainly obstructs the propagation of large-depth reversed sound rays, and has little impact on the normal mode waves on the surface, resulting in the main component of normal waves in the long-distance received sound signal, which is consistent with the results of the experimental received sound signal. The difference from the results in figure 28 is that the depth of the seamount on the propagation path from ICE08 station to NS station is relatively deep, and the seamount has a relatively small blocking effect on the low-frequency part of the normal mode, resulting in a relatively stronger energy of the first normal mode wave on this acoustic propagation path, and this can be seen from the experimental received acoustic signal in figure 32.

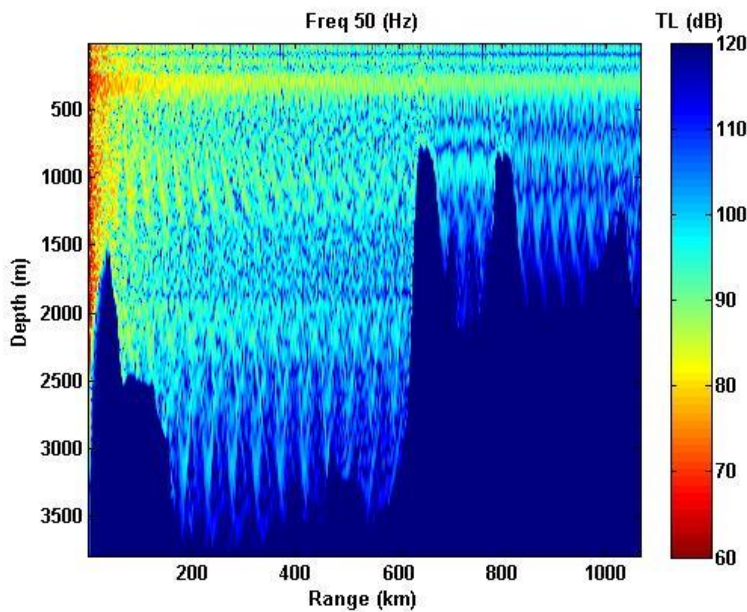

Figure 33 The calculation result of two-dimensional sound transmission loss from ICE08 station to NS station using the parabolic equation sound field model Ram

For the sound propagation problem from ICE08 station to NS station, due to the propagation distance of the sound signal being too far, the sound source station is located in the central ice zone at latitude of about 84.5 °N, and the receiving station is located in the Chukchi Plateau at latitude of about 75 °N, the SSPs of stations near the propagation path can be obtained using Figures 12 and 13, as shown in Figure 34. From the figure, it can be seen that, like the sound propagation path from ICE05 station to NS station, the SSP undergoes significant changes on the propagation path, from a single channel SSP to a dual-channel SSP.

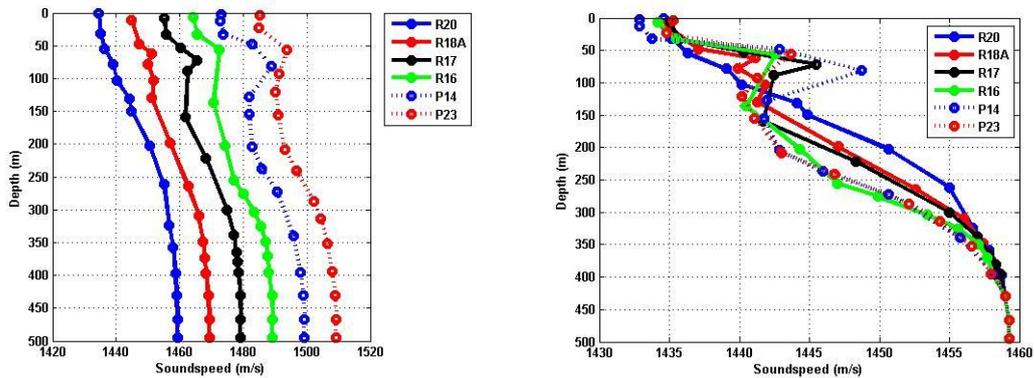

Figure 34 SSPs of six nearby stations on the sound propagation path from ICE08 sound source station to NS receiving station

In order to analyze the impact of changes in SSPs on long-distance propagation of sound signals, this article uses two types of SSPs for simulation. The first is to assume that the SSP remains unchanged horizontally and is fixed as a single-channel SSP in the central ice region of the Arctic where the sound source is located. The second is to assume that the SSP changes horizontally and interpolate the six SSPs in Figure 34 to obtain a sound velocity field that varies with distance. Figure 35 shows the time-domain waveform and time-frequency analysis of the simulated acoustic signal in two different scenarios, as well as a comparison with the time-domain waveform of the experimental signal and the extracted dispersion curve. The differences between the simulated acoustic signals in the two scenarios are significant. Considering the horizontal variation of the SSP, the duration of the sound signal is longer because the average sound velocity of the surface water in the dual-channel SSP is significantly lower than that in the single-channel SSP, which causes the high-frequency part of the normal mode to arrive later, thus making the duration of the sound signal longer. At the same time, it can be seen from the figure that the time-domain waveform and dispersion structure of the acoustic signal calculated considering the horizontal variation of the SSP are more consistent with the experimental results, indicating that the influence of the horizontal variation of the SSP needs to be considered in the study of long-distance sound propagation in the Arctic Ocean. In addition, the upper frequency limit of the normal mode of the simulated acoustic signal is significantly greater than that of the experimental acoustic signal. This is because the effect of sea ice cover is not considered in the acoustic signal simulation, resulting in the high-frequency part of the normal mode of the simulated acoustic signal not being attenuated, resulting in inconsistency between the two.

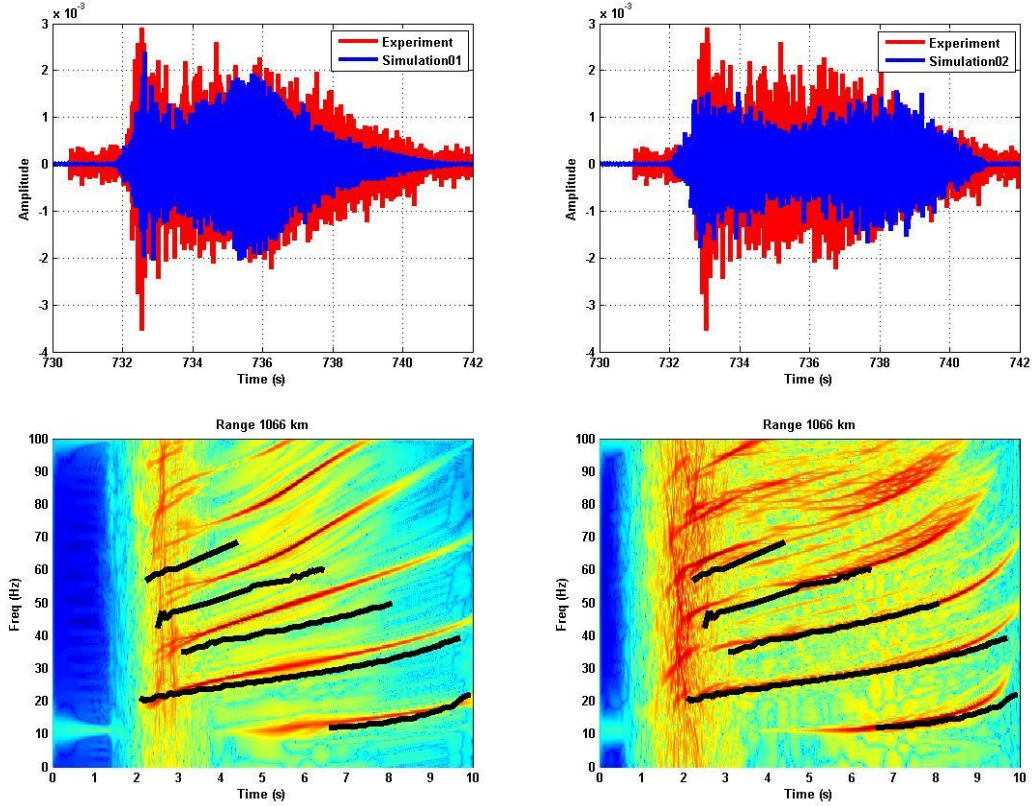

Figure 35 The time-domain waveform and time-frequency analysis of simulated acoustic signals from ICE08 station to NS station based on a horizontally invariant SSP and a horizontally varying SSP, as well as comparison with the time-domain waveform and extracted dispersion curve of the experimental measured signal

# VI. CONCLUSION

## A. Horizontal wavenumber estimation and warping transform of refractive normal modes under Arctic SSP

This article derives an approximate expression for the horizontal wavenumber of the Arctic surface waveguide based on the theory of normal modes, and further derives the warping transformation operator for the Arctic refractive normal modes. From simulation and experimental data, it can be seen that when the energy of the acoustic signal is mainly concentrated in the Arctic surface water with positive gradient SSP, the warping transformation operator of refractive normal modes can be used to separate various normal modes.

## B. Signal time-domain waveform and dispersion structure of Arctic low-frequency long-range sound propagation simulation

This article conducts simulation of low-frequency long-range sound propagation in the Arctic based on two types of measured typical Arctic SSP data, including simulation of two-dimensional sound propagation loss, simulation of normal wave eigenfunctions, simulation of acoustic signal time-domain waveforms, and simulation of acoustic signal dispersion structures. It analyzes the vertical distribution characteristics of eigenfunctions, explains the characteristics of acoustic signal time-domain waveforms and dispersion structures, and analyzes the impact of SSP on the characteristics of sound signals.

From the above simulation analysis, it can be concluded that under the typical SSP conditions of the central ice region in the Arctic, after long-distance propagation, the received sound signal is composed of sound rays that arrive in multiple paths and normal modes that arrive in multiple modes. The sound rays that arrive in multiple paths arrive earlier due to the higher sound velocity of the water passing through, while the normal modes arrive later due to the lower sound velocity of the surface water passing through.

Under the conditions of the dual-channel SSP in the Canadian Basin, the duration of the sound signal is longer than that under the conditions of the central ice zone SSP. This is because the surface average sound velocity of the dual-channel SSP is relatively smaller, resulting in a later arrival time of the normal mode, resulting in a longer duration of the sound signal.

By analyzing the eigenfunctions of normal waves under the two typical sound velocity profiles mentioned above, it can be concluded that the higher the frequency of normal waves, the more concentrated their energy is in the surface waveguide. Therefore, the upper frequency limit of normal modes is limited by the depth of the sound source and reception. If there is sea ice cover on the sea surface, the high-frequency part of normal modes is more attenuated by the ice-water interface. When the frequency of a normal mode is lower, the vertical coverage depth range of the eigenfunction of the normal mode is larger, so the lower limit of the frequency of the normal mode is mainly affected by the seabed.

In addition, under different SSP conditions, the coverage range of the eigenfunctions of the same normal mode at the same frequency is also different, resulting in different propagation speeds of the normal mode.

In summary, this article utilizes the vertical distribution of normal mode eigenfunctions under different SSP conditions as a function of mode number and frequency, explaining the characteristics of the time-domain waveform and dispersion structure of the received sound signal.

## C. Signal time-domain waveform and dispersion structure of Arctic low-frequency long-range sound propagation experiment

This article uses data from a low-frequency long-range acoustic propagation experiment under Arctic ice to study the time-domain waveform and dispersion structure of received acoustic signals, as well as their relationship with marine environmental parameters.

Due to the approximate absence of sea ice covering seawater on the sound propagation paths from S11 and S22 stations to NS station, these two sound propagation paths only consider the effects of changes in seabed topography and changes in seawater SSPs on sound propagation. The sound propagation signal from station S11 to station NS has a distinct 11 modes, and a distinct 6 modes can be separated from the signal. On this path, the influence of the seabed terrain is relatively small, and the signal can observe the multi-path sound rays with large depth reversal and early arrival. Due to the short propagation distance on this path, the horizontal variation of the SSP is relatively small and has little impact on sound propagation. The simulated time-domain waveform and dispersion structure are consistent with the measured results. At the same time, the frequency dispersion structure of the measured sound signal is basically consistent with the simulation results, including the upper and lower frequency limits of each normal mode.

For the sound propagation situation from station S22 to station NS, the seabed terrain changes strongly from a shallow water depth of 400m to a deep sea depth of about 2000m. The measured received acoustic signal has a distinct 5 normal modes. Due to the influence of seabed topography, each mode has a clear lower frequency limit. Due to the lack of sea ice cover on the sea surface, the upper limit frequency of normal mode waves is only related to the depth of the sound source and the depth of reception. At the same time, due to the influence of the seabed terrain, there is no significant multi-path arrival sound ray in the received sound signal due to the inability to have a deep reversed sound ray. Using warping transform can effectively extract the 5 normal modes from the acoustic signal. Due to the sound propagation path having a SSP that varies from west to east, that is, from a non dual-channel SSP to a dual-channel SSP, after considering this factor, the simulated signal dispersion structure is basically consistent with the measured results, and the upper and lower frequency limits of the normal mode wave are also basically consistent with the experimental results. This sound propagation path well demonstrates the sound propagation from shallow water to deep sea under the conditions of the Arctic SSP.

For the sound propagation path from ICE05 station to NS station, the seabed terrain of this sound propagation path changes from the Canadian Basin to the Chukchi Plateau, and the SSP changes from no dual-channel SSP in the central ice zone to a dual-channel SSP, while this sound propagation path has a clear sea ice cover. The obvious 4 normal modes can be observed from the measured sound signal, and these four normal modes can be separated from the signal using warping transformation. Due to the existence of seamounts in the propagation path, the energy of the first normal wave is relatively weak. After considering the changes in seabed topography and SSP, the dispersion structure of the simulated sound signal is basically consistent with the measured results. However, due to the lack of consideration of the impact of sea ice cover in the simulation, the time-domain waveform and upper frequency limit of the normal mode of the simulated acoustic signal are inconsistent with the experimental results.

For the sound propagation paths from ICE08 station and ICE06 station to NS station, the situation is basically the same as that from ICE05 station to NS station. Except for the fact that the shallowest depth of the first two sound propagation paths is greater than the shallowest depth of the last one, the first normal mode of the first two sound propagation paths is relatively less affected by the seabed terrain, and the obvious first normal mode can be observed from the measured signals. Based on the warping transform, the first five normal modes can be extracted from the received acoustic signal. After considering the changes in seabed topography and the horizontal variation of SSP, the dispersion structure of the simulated sound signal is basically consistent with the experimental results. However, due to the lack of consideration of acoustic signal attenuation caused by sea ice cover in the simulation, there are certain differences between the time-domain waveform and the upper frequency limit of the normal mode of the simulated acoustic signal and the experimental results. In future work, this issue will be addressed by introducing acoustic attenuation at the ice water interface.

This article analyzes the time-domain waveform and dispersion structure characteristics under typical Arctic SSP conditions through several typical examples of sound propagation signals. At the same time, it analyzes the effects of changes in seabed topography, sea water SSP, and sea ice cover on sound signal characteristics.

## D. Separation of refractive normal modes based on a single hydrophone

This article utilizes experimental data of Arctic sound propagation to achieve normal mode separation based on a single hydrophone in the unique marine environment in the Arctic. Compared to previous methods of using large aperture synchronous vertical arrays for normal mode separation, the method adopted in this article has the advantages of low cost and easy deployment. In addition, previous warping transformations were mainly applicable to shallow sea areas, and this article implemented the normal mode warping transformation in the deep Arctic sea.

## E. Incomplete research content

A problem in this article is that the acoustic attenuation caused by sea ice cover was not considered in the simulation of received acoustic signals, which affects the consistency between the simulated and experimental acoustic signals. Another issue in this article is that in the Arctic sound propagation experiment, accurate propagation distance cannot be obtained due to the lack of precise measurement of sound source and receiving positions, the propagation time of sound signals, the seabed topography and sound velocity profile along the sound propagation path, and the use of up looking sonar to measure sea ice parameters. In future theoretical and experimental research work, the issues raised above will be further improved. Additionally, this article did not conduct an analysis of the intensity of received acoustic signals. The next step of work will be to conduct research on the propagation loss of various orders of normal modes, and analyze the factors that affect the propagation loss of normal modes.

This article utilizes a long-range acoustic propagation experiment under the Arctic ice based on high source level broadband pulse sound source signals to deeply reveal the mechanism of the influence of seabed topography, seawater SSP, and sea ice cover on acoustic propagation in the Arctic marine environment. It has guiding significance for remote acoustic detection and positioning under the Arctic ice, communication navigation, and acoustic monitoring of large-scale changes in the Arctic marine environment.

## ACKNOWLEDEMENTS

This research was supported in part by the National Key Research and Development Program of China under Grant, in part by the National Natural Science Foundation of China under Grant, in part by the Scientific Research Foundation of Third Institute of Oceanography, State Oceanic Administration under Grant, and in part by the Opening Foundation of State key laboratory of acoustics under Grant. Special thanks to all the staff involved in the deployment and recycling of equipment during the experiment.

## REFERENCES

1. B. M. Buck and C. R. Green, Arctic deep-water propagation measurements, The Journal of the Acoustical Society of America, 1964, 36, 1526-1533
2. Finn B. Jensen, William A. Kuperman, Michael B. Porter, Henrik Schmidt, Computational Ocean Acoustics, Second Edition, Springer
3. Cockrell K L, Schmidt H, 2011 J. Acoust. Soc. Am. 130 72
4. Bender C M, Orszag S A 1978 Advanced Mathematical Methods for Scientists and Engineers (New York: McGraw-Hill) p276
5. Kelly A. Keen, Bruce J. Thayre, John A. Hildebrand, Sean M. Wiggins, Seismic airgun sound propagation in Arctic Ocean waveguides, Deep-Sea Research Part I, 2018.09.003
6. Megan S. Ballard, Mohsen Badiey, Jason D. Sagers, et al. Temporal and spatial dependence of a yearlong record of sound propagation from the Canada Basin to the Chukchi Shelf, The Journal of the Acoustical Society of America, 148, 2020, 1663-1680
7. Arthur B. Baggeroer and Jon M. Collis, Transmission loss for the Beaufort Lens and the critica frequency for mode propagation during ICEX-18, 151, 2022, 2760-2772
8. Timothy F. Duda, Weifeng Gordon Zhang, and Ying-Tsong Lin, Effects of Pacific Summer Water layer variations and ice cover on Beaufort Sea underwater sound ducting, The Journal of the Acoustical Society of America, 2021, 149, 2117-2136
9. Kevin Lepage and Henrik Schmidt, Modeling of low-frequency transmission loss in the central Arctic, The Journal of the Acoustical Society of America, 1994, 96, 3, 1783-1795
10. Mural Kucukosmanoglu, John A. Colosi, Peter F. Worcester, Mattew A. Dzieciuch, and Daniel J. Torres, Observations of sound-speed fluctuations in the Beaufort Sea from summer 2016 to summer 2017, The Journal of the Acoustical Society of America, 2021, 149, 1536-1548
11. Aaron Thode, Katherine H. Kim, Charles R. Greene, et al. Long range transmission loss of broadband seismic pulses in the Arctic under ice-free conditions, The Journal of the Acoustical Society of America, 2010, EL181-187, 128
12. T. C. Yang, Dispersion and ranging of transient signals in the Arctic Ocean, The Journal of the Acoustical Society of America, 1984, 76(1), 262-273
13. T. C. Yang, A method of range and depth estimation by modal decomposition, The Journal of the Acoustical Society of America, 1987, 82, 1736-1745
14. T. C. Yang, Effectiveness of mode filtering: A comparison of matched-field and matched-mode processing, The Journal of the Acoustical Society of America, 1990, 87, 2072-2083
15. Alexander N. Gavrilov, Peter N. Mikhalevsky, Low-frequency acoustic propagation loss in the Arctic Ocean: results of the Arctic Climate Observations using Underwater Sound experiment, The Journal of the Acoustical Society of America, 2004
16. W. A. Kuperman, Henrik Schmidt, Self-consistent perturbation approach to rough surface scattering in stratified elastic media, The Journal of the Acoustical Society of America, 1989, 86, 4, 1511-1522
17. Peter N. Mikhalevsky, and Alexander N. Gavrilov, Acoustic thermometry in the Arctic Ocean, Polar Research, 2001, 185-192
18. Hanne Sagen, Matthew Dzieciuch, et al, The Coordinated Arctic Acoustic Thermometry Experiment – CAATEX
19. Baraniuk R G, Jones D L, Unitary equivalence: A new twist on signal processing, IEEE Trans. Sign. Process, 1995, 43(10):2269-2282
20. Le Touze G, Nicolas B, Mars J I, el al, Matched representations and filters for guided waves, IEEE Trans. Sign. Process, 2009, 57(5), 1783-1795
21. Bonnel J, Chapman N R, Geoacoustic inversion in a dispersive waveguide using warping operators, Journal of the Acoustical Society of America, 2011, 130(2), EL101-107
22. Bonne Julien, Aaron Thode, Dana Wright and Ross Chapman, Nonlinear time-warping made simple: A step-by-step tutorial on underwater acoustic modal separation with a single hydrophone, 2020, 1897, 147
23. Niu haiqiang, Renhe Zhang, and Zhenglin Li, Theoretical analysis of warping operators for non-ideal shallow water waveguides, Journal of the Acoustical Society of America, 2014, 136, 53,
24. Qi yubo, Shihong Zhou, renhe zhang, yun ren, a waveguide-invariant-based

    warping operator and its application to passive source range estimation, Journal of Computational Acoustics, 2015, 23: 1550003-1-1550003-19
25. Yubo Qi, Shihong Zhou, Renhe Zhang, Bo Zhang, Yun Ren, Modal characteristic frequency in a range-dependent shallow-water waveguide and its application to passive source range estimation, Acta Phys. Sin., 2014, 63(4), 044303-1-9(in Chinese)
26. Yubo Qi, Shihong Zhou, Renhe Zhang, Yun Ren, A passive source ranging method using the waveguide-invariant-warping operator, Acta Phys. Sin., 2015, 64(7), 074301-1-6(in Chinese)
27. Julien Bonnel, Aaron M. Thode, Susanna B. Blackdwell and Kathering Kim, A. Michael Macrander, Range estimation of bowhead whale (Balaena mysticetus) calls in the Arctic using a single hydrophone, Journal of the Acoustical Society of America, 2014, 136, 145
28. Julien Bonnel, Stan dosso, N. Ross chapman, Bayesian geoacoustic inversion of single hydrophone light bulb data using warping dispersion analysis, Journal of the Acoustical Society of America, 2013, 134, 120
29. Y. B. Qi, Zhou Shi-Hong, Zhang Ren-He, Warping transform of the refractive normal mode in a shallow water waveguide, Acta Physica Sinica, 2016
30. Michael D. Collins, User's Guide for RAM Version 1.0 and 1.0p
31. Michael B. Porter, The Kraken Normal Mode Program (DRAFT)
32. Deming Zhang, Zhenglin Li, Renhe Zhang, Inversion for the bottom geoacoustic parameters based on adaptive time-frequency analysis, ACTA ACUSTICA, 2005, 30(5), 415-419(in Chinese)